\documentclass[journal, 10pt]{IEEEtran}
\IEEEoverridecommandlockouts
\usepackage{amsmath,graphicx}
\usepackage{amsmath,amssymb,amsfonts}
\usepackage{cite}
\usepackage{url}
\usepackage{bm}
\usepackage{comment}

\usepackage[usenames,dvipsnames]{color}
\usepackage{subcaption}
\usepackage{makecell}
\usepackage{algpseudocode}
\usepackage{color}

\usepackage{tabularx}     
\usepackage{textcomp}     
\usepackage{threeparttable} 
\usepackage{nth}          
\usepackage{gensymb}      

\usepackage{mathtools}    
\usepackage{amsfonts}
\usepackage{bm}
\usepackage{verbatim}

\usepackage{stfloats}
\usepackage{setspace}
\usepackage{caption, soul}

\usepackage{todonotes}
\usepackage{bbm}

\usepackage[linesnumbered,vlined,ruled]{algorithm2e}

\newenvironment{algocolor}{%
   \setlength{\parindent}{0pt}
   \itshape
    
}{}

\title{A Deep Joint Source-Channel Coding Scheme \\ for Hybrid Mobile Multi-hop Networks}
\author{Chenghong Bian, \IEEEmembership{Graduate Student Member,~IEEE}, Yulin Shao, \IEEEmembership{Member,~IEEE}, Deniz G{\"u}nd{\"u}z,\IEEEmembership{Fellow,~IEEE}
\thanks{C. Bian and D. G{\"u}nd{\"u}z are with the Department of Electrical and Electronic Engineering, Imperial College London, London SW7 2AZ, U.K. (e-mails: \{c.bian22, d.gunduz\}@imperial.ac.uk).}
\thanks{Y. Shao is with the State Key Laboratory of Internet of Things for Smart City and the Department of Electrical and Computer Engineering, University of Macau, Macau S.A.R (e-mail: ylshao@um.edu.mo) \textit{(Corresponding author: Yulin Shao)}.}
\thanks{This work received funding in part from the UKRI for the projects AI-R (ERC Consolidator Grant, EP/X030806/1) and INFORMED-AI (EP/Y028732/1), and in part from the SNS JU project 6G-GOALS under the EU's Horizon program (Grant Agreement No. 101139232). The work of Y. Shao was supported by the Science and Technology Development Fund, Macao (Project no.: 0068/2023/RIB3 and 0062/2024/RIA1), and the Multi-Year Research Grant, University of Macau (Project no.: MYRG-CRG2024-00011-IOTSC).}
\thanks{For the purpose of open access, the authors have applied a Creative Commons Attribution (CC BY) license to any author accepted manuscript version arising.}
\thanks{This paper was presented in part at the IEEE International Conference on Communications, 2024 \cite{hybrid_jscc}.}
}

\begin{document}

\maketitle
\thispagestyle{empty}
\begin{abstract}
Efficient data transmission across mobile multi-hop networks that connect edge devices to core servers presents significant challenges, particularly due to the variability in link qualities between wireless and wired segments.  This variability necessitates a robust transmission scheme that transcends the limitations of existing deep joint source-channel coding (DeepJSCC) strategies, which often struggle at the intersection of analog and digital methods. Addressing this need, this paper introduces a novel hybrid DeepJSCC framework, h-DJSCC, tailored for effective image transmission from edge devices through a network architecture that includes initial wireless transmission followed by multiple wired hops. Our approach harnesses the strengths of DeepJSCC for the initial, variable-quality wireless link to avoid the cliff effect inherent in purely digital schemes. For the subsequent wired hops, which feature more stable and high-capacity connections, we implement digital compression and forwarding techniques to prevent noise accumulation. This dual-mode strategy is adaptable even in scenarios with limited knowledge of the image distribution, enhancing the framework's robustness and utility. Extensive numerical simulations demonstrate that our hybrid solution outperforms traditional fully digital approaches by effectively managing transitions between different network segments and optimizing for variable signal-to-noise ratios (SNRs). We also introduce a fully adaptive h-DJSCC architecture {with both SNR-adaptive (SA) and rate-adaptive (RA) modules} capable of adjusting to different network conditions and achieving diverse rate-distortion objectives, thereby reducing the memory requirements on network nodes.
\end{abstract}


\section{Introduction}
\label{sec:intro}

\begin{figure}[!t]
\centering
\includegraphics[width=\columnwidth]{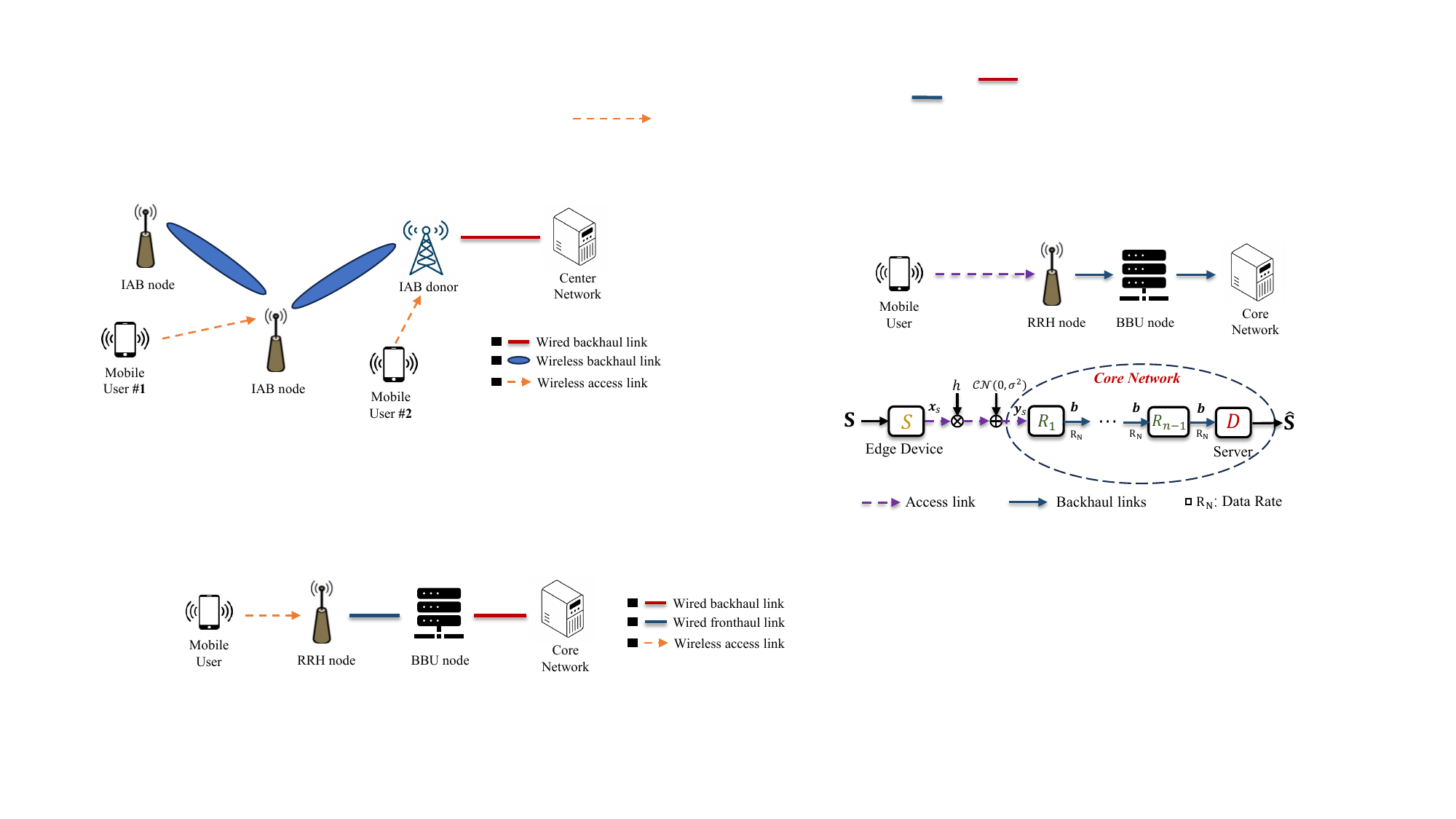}
\caption{{Illustration of efficient data transmission over a hybrid multi-hop network like C-RAN (top). An edge device $\mathrm{S}$ transmits data to an edge server $\mathrm{D}$ via a sequence of relay nodes ${\mathrm{R}1, \ldots, \mathrm{R}_{n-1}}$ within the core network. The variable $h$ represents the fading coefficient of the wireless link.}}
  \label{fig:fig_system}
\end{figure}

Multi-hop relay networks are essential for enhancing the coverage of wireless communication systems, enabling seamless connectivity across widely dispersed areas. These networks support the transmission of messages through a sequence of relay nodes, substantially increasing the reach and reliability of wireless communications by overcoming the limitations imposed by physical distance and challenging terrains.
{
A notable application of this principle is found in cloud radio access networks (C-RANs), where data transmission begins at the mobile edge. In this setup, devices initially connect to a nearby remote radio head (RRH) through a variable-quality wireless access link. From there, the data travels through multiple wired fronthaul and backhaul links, which are processed by the baseband unit (BBU) before reaching the core network server, as depicted in Fig.~\ref{fig:fig_system}. These wired links, typically characterized by fixed rate constraints, offer stable and high-capacity connections that are critical for efficient and reliable data management.
}


{
Building on the concept of hybrid multi-hop relays in C-RANs, this paper explores the end-to-end quality of source transmission over such networks, by adapting and extending the concept of joint source-channel coding (JSCC) \cite{gündüz2024jointsourcechannelcodingfundamentals} towards practical deployment in realistic multi-hop network scenarios.
Recent advancements in deep learning (DL) have driven remarkable progress in the design of JSCC schemes tailored for wireless channels.}
Originally proposed in \cite{deepjscc}, the DeepJSCC methodology has been recognized for its ability to surpass traditional digital communication benchmarks that combine state-of-the-art image compression techniques with near-optimal channel codes. Moreover, it achieves a graceful performance degradation in scenarios of deteriorating channel conditions. In \cite{semantic_video_seu, bupt_video, semantic_speech1, sept}, the DeepJSCC framework is extended to video, speech and point cloud sources, highlighting it as an effective tool for a large variety of applications. More recently, DeepJSCC techniques have also been extended to multi-user scenarios, where \cite{bian2024processandforward,semantic_relay_speech, IOTJ_2024, mac_jscc, broadcast} show the superior performance of the DeepJSCC in the three node cooperative relay, the multiple access, and broadcast channels, respectively. This paper aims to explore the potential of DeepJSCC in optimizing data transmission from edge devices through a hybrid network of wireless and wired links, ensuring robust and efficient data delivery to edge servers. The most related works are \cite{semantic_multi-hop1, zju_multihop}, where the authors investigate DeepJSCC over multi-hop networks. In particular, \cite{semantic_multi-hop1} introduces a term named semantic similarity and optimizes the neural networks to preserve the similarity over the hops, whereas \cite{zju_multihop} proposes a recursive training methodology to mitigate the noise accumulation problem. Though some performance improvements are attained compared with the naively trained scheme, \cite{semantic_multi-hop1, zju_multihop} fail to fit into the considered scenario with wired links. 

A key factor underpinning the success of DeepJSCC is its reliance on discrete-time analog transmission (DTAT) \cite{DTAT}, which introduces a greater degree of flexibility for channel symbols and ensures that the reception quality at the receiver is closely tied to fluctuations in channel quality. This refined strategy is in sharp contrast to conventional digital transmission methods, which are characterized by a binary outcome: successful decoding or complete data loss, the latter leads to a precipitous performance drop-off known as the \textit{cliff effect}.
However, while the DTAT approach offers distinct advantages in scenarios involving direct, single-hop transmissions, it poses significant challenges within the context of networks involving multiple transmission mediums. 
{In this architecture, data transmitted from the mobile user to the RRH node via wireless access links and then relayed through a wired C-RAN network to the center network, introduces an interplay between different transmission characteristics. The initial wireless access link, susceptible to channel impairments such as fading and noise, and the subsequent wired hops with fixed rate constraints, require careful management to maintain signal integrity.}
Here, the signal distortion introduced in the wireless segment and potential bottlenecks in wired segments can lead to a degradation of signal quality -- a problem less pronounced in digital systems where data can be regenerated at each node. Consequently, the performance of DeepJSCC tends to decline as the network complexity increases, calling for innovative solutions tailored for hybrid multi-hop network architectures.


In this paper, we introduces a novel DeepJSCC framework, h-DJSCC, tailored for efficient image transmission across hybrid mobile multi-hop networks.
The cornerstone of h-DJSCC is the strategic use of DeepJSCC for the first hop (i.e., the access link), capitalizing on its ability to offer variable quality transmission. This choice effectively sidesteps the notorious cliff effect associated with digital schemes, ensuring a more resilient transmission over the wireless hop.
For the subsequent hops through a wired network with fixed rate constraints, h-DJSCC transitions to a digital transmission mode. This dual-phase approach marries the best of both worlds: it leverages the adaptability of DeepJSCC to mitigate variations in the wireless channel conditions while employing digital transmission in the later stages to overcome bandwidth limitations and ensure consistent, error-free data transmission. {This strategy offers a nuanced balance, providing a robust solution for high-quality image transmission over practical wireless networks with reliable backhaul connections.}

Our main contributions are summarized as follows.
\begin{itemize}
    \item We introduce h-DJSCC, a novel DeepJSCC framework designed for multi-hop image transmission from mobile edge devices to the core network. This approach leverages a hybrid communication strategy that begins with an analog-based DeepJSCC scheme for the initial wireless link under fluctuating channel conditions, followed by digital coding and modulation schemes for the subsequent wired hops with fixed rate constraints. A neural network-based compression module is deployed at the access point to efficiently convert the received analog DeepJSCC codeword into a digital bit sequence, facilitating flexible trade-offs between transmission latency and reconstruction quality to accommodate diverse network demands. {We further show the effectiveness of the h-DJSCC in a scenario where two mobile users communicate with each other via the C-RAN network.}
    \item We explore the adaptability of the h-DJSCC framework to scenarios where the access point performs oblivious relaying -- processing signals based only on past received signals without knowledge of the transmitted image distribution. This aspect underscores the robustness of the h-DJSCC framework, showing notable performance improvements over traditional methods where relays directly quantize the received signals into bit sequences.
    \item We propose a fully adaptive h-DJSCC scheme suitable for scenarios where the initial wireless link experiences fluctuations. This scheme is robust across both additive white Gaussian noise (AWGN) and Rayleigh fading channels, demonstrating the model's capability to adjust to various signal-to-noise ratio (SNR) levels and achieve multiple rate-distortion (R-D) points. The fully adaptive h-DJSCC scheme is carefully initialized to avoid succumbing to sub-optimal solutions.
    \item {We further endow h-DJSCC with adaptibility to the available bit-rate of the backhaul links. This allows h-DJSCC to transmit only the features that can be conveyed through the wired links, and hence, increase their fidelity. To achieve this, a rate-adaptive (RA) module is proposed, which enables flexible feature adjustment according to different bit budgets for better performance.}
    \item Through comprehensive numerical experiments, we validate the h-DJSCC framework's superiority over traditional fully digital approaches across different channel conditions and datasets. Our results underscore the fully adaptive h-DJSCC framework's enhanced R-D performance, confirming its practical effectiveness and adaptability to diverse transmission scenarios. {The complexity of the proposed h-DJSCC scheme is further provided to show its potential for practical deployment.}
\end{itemize}

{\it Notations:} Throughout the paper, scalars are denoted by normal-face letters (e.g., $x$),  while random variables are denoted by uppercase letters (e.g., $X$). Matrices and vectors are denoted by bold {upper} and {lower} case letters (e.g., $\bm{X}$ and $\bm{x}$), respectively. {$L_{\bm{b}}$ represents the length of vector $\bm{b}$.}  {A set is denoted by $\mathbb{S}$}. Transpose and conjugate operators are denoted by $(\cdot)^\top$ and $(\cdot)^*$, respectively.

{
\section{Preliminary}\label{sec:preliminary}
This section offers a concise overview of learned image compression and the SNR-adaptive module, which form the foundation for the development of the proposed h-DJSCC scheme.

\begin{figure}[!t]
\centering
\includegraphics[width=0.8\columnwidth]{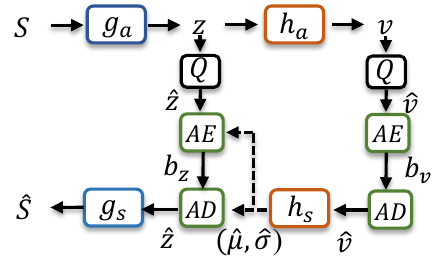}
\caption{{Illustration of the compression and decompression modules, $f_c(\cdot)$ and $g_c(\cdot)$, respectively. In particular, $g_a(\cdot), h_a(\cdot)$ are the non-linear transform blocks to generate the latent vectors $\bm{z, v}$, whereas $g_s(\cdot), h_s(\cdot)$ are adopted to facilitate the synthesis of $\hat{\bm{S}}, \hat{\bm{z}}$. The predicted tensors $\hat{\bm{\mu}}, \hat{\bm{\sigma}}$ are used for the arithmetic encoding and decoding of $\hat{\bm{z}}$ and $\bm{b}_z$.}}
\label{fig:fig_hyperprior}
\end{figure}

\subsection{Learned Image Compression}\label{sec:pre_img_comp}
Learned image compression leverages neural networks to optimize the image compression process, often outperforming traditional methods by jointly learning efficient representations and entropy models. These models are designed to minimize a combination of distortion and bitrate, resulting in variable-rate compression schemes that can be trained to operate on a desired point on the rate-distortion trade-off curve.

The processing of the neural image compression model \cite{balle2018variational} is summarized as follows. As illustration is given in Fig. \ref{fig:fig_hyperprior}.

Let $\bm{S} \in \mathbb{R}^{C\times H \times W}$ denote the input image,  where $C$, $H$, $W$ are the number of color channels, height and width of the image, respectively.  The encoder first uses non-linear transformation blocks, denoted by $g_a(\cdot)$, comprised of 2D convolutional neural networks (CNNs) with $C_z$ output channels to generate a latent tensor $\bm{z}\in \mathbb{R}^{C_z\times H/4 \times W/4}$, which is fed to the hyper-latent encoder, $h_a(\cdot)$, to generate the hyper-latent $\bm{v}\in \mathbb{R}^{C_v\times H/16 \times W/16}$:
\begin{equation}
    \bm{z} = g_a({\bm{S}}); \quad \bm{v} = h_a(\bm{z}).
    \label{equ:z_v}
\end{equation}
The hyper-latent $\bm{v}$ is introduced to remove the dependency between the elements of $\bm{z}$, such that given $\bm{v}$, the probability density function (pdf) of $\bm{z}$ will be a product of its marginals. 

In the deployment phase, we round each element of the two latent vectors, ${z}_i$ and ${v}_i$, to the nearest integers, $\hat{{z}}_i$ and $\hat{{v}}_i$:
\begin{equation}
    \hat{{z}}_i, \hat{{v}}_i = Q(z_i), Q(v_i),
    \label{equ:quantize}
\end{equation}
where $Q(\cdot)$ denotes the quantization operation.
We then arithmetically encode the quantized vectors according to their own probability distributions. In the training phase, however, quantization does not allow back propagation. Thus, during training, we use $\tilde{\bm{z}}, \tilde{\bm{v}}$ to replace the quantized vectors $\hat{\bm{z}}, \hat{\bm{v}}$, where the elements $\tilde{{z}}_i, \tilde{{v}}_i$ are obtained by adding uniform noise, $\mathcal{U}(-\frac{1}{2}, \frac{1}{2})$ (to model the quantization noise) to ${z}_i, {v}_i$. A hyper latent decoder, $h_s(\cdot)$, with two upsampling layers takes $\tilde{\bm{v}}$ as input, and outputs two tensors\footnote{{Note that the $\tilde{\bm{\sigma}}$ here is different from the $\sigma^2$ in \eqref{equ:y1} which represents the power of the channel noise.}} $\tilde{\bm{\mu}}$ and $\tilde{\bm{\sigma}}$ with the same dimension as $\tilde{\bm{z}}$. Then, we model the pdf of $\tilde{\bm{z}}$ as:
\begin{align}
    p(\tilde{\bm{z}}|\tilde{\bm{v}}) = &\prod_i \left(\mathcal{N}(\tilde{\mu}_i, \tilde{\sigma}^2_i)*\mathcal{U}(-\frac{1}{2}, \frac{1}{2}) \right)(\tilde{z}_i), \notag \\
    \widetilde{\bm{\mu}}, \widetilde{\bm{\sigma}} &= h_s(\tilde{\bm{v}}, \bm{\theta}),
\label{equ:probz}
\end{align}
where $\bm{\theta}$ denotes the parameters of $h_s(\cdot)$ and $*$ represents convolution.
For the hyper latent $\tilde{\bm{v}}$, we consider modeling it also using a fully factorized model as:
\begin{align}
    p(\tilde{\bm{v}}|\bm{\phi}) = \prod_i \left(p_{\tilde{v}_i|\phi_i}(\phi_i)*\mathcal{U}(-\frac{1}{2}, \frac{1}{2}) \right)(\tilde{v}_i),
\label{equ:probv}
\end{align}
where $\bm{\phi}$ denotes a collection of parameters to parameterize the univariate distribution. 

Once the models are trained and deployed, we use an arithmetic encoder (AE) to encode the rounded $\hat{\bm{z}}, \hat{\bm{v}}$ to the bit sequence $\bm{b} = (\bm{b}_z, \bm{b}_v)$ according to the optimized pdfs in \eqref{equ:probz} and \eqref{equ:probv} by simply replacing $\tilde{\bm{z}}, \tilde{\bm{v}}$ by $\hat{\bm{z}}, \hat{\bm{v}}$. 
The image decompressor employs an arithmetic decoder ($AD$) to first generate the rounded latent vector $\bm{\hat{v}}$, which is then fed to $h_s(\cdot)$ to obtain $\hat{\bm{\mu}}, \hat{\bm{\sigma}}$. Then, another $AD$ takes the mean and variance, $\hat{\bm{\mu}}, \hat{\bm{\sigma}}$ along with the sequence $\bm{b}_z$ to produce the rounded latent vector $\bm{\hat{z}}$. Finally, the decompressor, $g_c(\cdot)$ takes $\bm{\hat{z}}$ as input and outputs the reconstructed image $\hat{\bm{S}}$. 

\textbf{Loss function}.
To achieve different trade-off points on the rate-distortion curve, we introduce a variable $\lambda$ and the loss function is written as:
\begin{equation}
    \mathcal{L} = \lambda \|\bm{S} - \hat{\bm{S}}\|^2_F + I,
    \label{eq:loss_fun}
\end{equation}
where $I = I_z + I_v$ is the summation of the bit per pixel (bpp) for compressing $\tilde{\bm{z}}$ and $\tilde{\bm{v}}$, respectively, which is defined by the number of bits divided by the height $H$ and width $W$ of the image. Note that the notation $\tilde{\bm{z}}, \tilde{\bm{v}}$ indicate the training phase\footnote{{Definition of $I$ is the same for the deployment phase obtained by replacing $\tilde{\bm{z}}$ and $\tilde{\bm{v}}$ with $\hat{\bm{z}}$ and $\hat{\bm{v}}$.}} and the rate can be further expressed as:
\begin{align}
    I = \frac{1}{HW} \mathbb{E}_{\tilde{\bm{z}}, \tilde{\bm{v}} \sim q} \left[-\log_2(p_{\tilde{\bm{z}}|\tilde{\bm{v}}}(\tilde{\bm{z}}|\tilde{\bm{v}})) - \log_2(p_{\tilde{\bm{v}}}(\tilde{\bm{v}})) \right],
\label{equ:Izv}
\end{align}
where the first and second terms represent $I_z$ and $I_v$, respectively, while $q$ denotes the posterior of $\tilde{\bm{z}}, \tilde{\bm{v}}$ given the input image ${\bm{S}}$, which follows a uniform distribution centered at $g_a(\cdot)$ and $h_a(\cdot)$ output, ${\bm{z}}$ and ${\bm{v}}$:
\begin{align}
    q(\tilde{\bm{z}}, \tilde{\bm{v}}|\bm{{S}}) = \prod_i \mathcal{U}(&\tilde{{z}}_i|{z}_i-\frac{1}{2}, {z}_i+\frac{1}{2}) \notag \\
    &\mathcal{U}(\tilde{{v}}_i|{v}_i-\frac{1}{2}, {v}_i+\frac{1}{2}).
\label{equ:post}
\end{align}

\subsection{SNR-adaptive Module}\label{sec:pre_sa_module}
The SNR-adaptive (SA) module \cite{xu2021wireless} enables a single DeepJSCC model to achieve reconstruction performance comparable to that of individually trained models across a wide range of SNRs. The key idea is that the SA module allows the DeepJSCC encoder to dynamically adjust which features to transmit and at what level of fidelity, depending on the channel conditions.

To be more specific, we denote the input feature to the SA module as $\bm{Z}_l\in \mathbb{R}^{C_l\times H_l\times W_l}$, where $C_l, H_l, W_l$ refer to the number of channels, the height and the width of the feature, respectively. 
Then, the weight $\bm{w} \in \mathbb{R}^{C_l}$ is calculated according to $\bm{Z}_l$ and the SNR value, denoted as $\eta$, and is further multiplied with the input feature in a channel-wise manner to obtain the output $\bm{Z}_{l, SA}\in \mathbb{R}^{C_l\times H_l\times W_l}$:
\begin{align}
    \bm{w} &= \text{MLP}\left(\frac{1}{H_l W_l}\sum_{i,j} \bm{Z}_l[:,i,j],\eta\right), \notag \\
    \bm{Z}_{l, SA}[k] &= {w}_k  \bm{Z}_l[k], \quad k \in [1, C_l],
    \label{equ:snr_adapt}
\end{align}
where $\text{MLP}$ denotes the multi-layer perceptron.
Note that different channels of $\bm{Z}_l$ may contain different features, and by multiplying the SNR-aware weight, $\bm{w}$, with $\bm{Z}_l$, a different balance between the information content of the conveyed features and level of protection against channel noise is achieved for different $\eta$ values. }

\section{Problem Formulation}\label{sec:problem}

\subsection{System Model}\label{sec:system_model}
We consider a multi-hop network, where a mobile edge user $\mathrm{S}$ aims to deliver an image to a destination node $\mathrm{D}$ with the aid of $(n-1)$ relay nodes $\{\mathrm{R}_1, \ldots, \mathrm{R}_{n-1}\}$ residing in the core network, as depicted in Fig.~\ref{fig:fig_system}.
The backhaul links within the core network are more reliable than the access link between $\mathrm{S}$ and $\mathrm{R}_1$. 
In particular, we assume the backhaul links are able to deliver information errorlessly and each link is associated with an achievable rate, $R_i$. Without loss of generalizability, we assume all the nodes are of the same quality, with achievable rates $R_1 = \ldots = R_n = R_N$. As illustrated in Fig. \ref{fig:fig_system}, the transmission over the backhaul links within the core network can be abstracted as a single hop with an achievable rate $R_N$.\footnote{This is valid for full-duplex relays. The achievable rate reduces to $\frac{R_N}{n}$ if half-duplex relays are considered.}
Initially, we base our analysis on the premise that the wireless channel in the first hop operates as AWGN channel. This assumption lays the groundwork for our foundational model, which we will subsequently extend to encompass Rayleigh fading channels in the later sections.

\begin{figure}[!t]
\centering
\includegraphics[width=0.9\columnwidth]{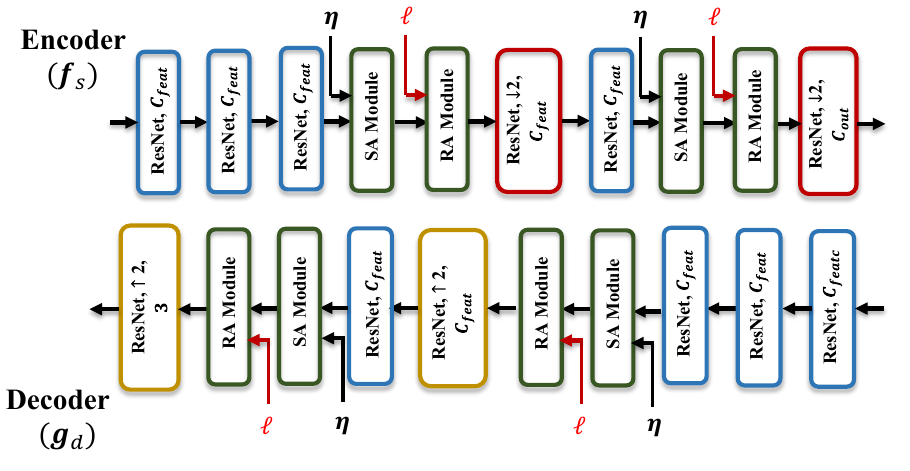}
\caption{{Neural network architecture for the proposed DeepJSCC encoder $f_s(\cdot)$ and decoder $g_d(\cdot)$ with SNR-adaptive (SA) and the rate-adaptive (RA) modules introduced in Section \ref{sec:full_adapt_jsc}.}}
\label{fig:fig_NN}
\end{figure}

The edge user $\mathrm{S}$ encodes image $\bm{S} \in \mathbb{R}^{C\times H \times W}$ to channel input vector $\bm{x}_s \in \mathbb{C}^k$ using an encoding function\footnote{{We note that $k$ refers to the expected number of channel uses if variable length source encoder is adopted.}} $f_s(\cdot): \mathbb{R}^{C\times H \times W} \rightarrow \mathbb{C}^k$, where $C$, $H$, $W$ denote the number of color channels, the height and the width of the image, respectively. The channel input is subject to a power constraint:
\begin{equation}
\frac{1}{k}\|\bm{x}_s \|^2 \leq 1.
\label{equ:pn}
\end{equation}

The signal received at the first relay node $\mathrm{R}_1$ (i.e., the access point of the core network) can be expressed as
\begin{equation}
\bm{y}_{1} = \bm{x}_s + \bm{w}_{s},
\label{equ:y1}
\end{equation}
where $\bm{w}_s \sim \mathcal{CN}(0, \sigma^2 \bm{I}_{k})$. The SNR of the first hop is defined as $\eta \triangleq \frac{1}{\sigma^2}$.

Within the core network, the relays can adopt different processing strategies, such as the fully digital scheme, the naive quantization scheme, and the proposed h-DJSCC scheme detailed later. Different strategies correspond to different processing at the first relay, $\mathrm{R}_1$, which is denoted by $f_{\mathrm{R}_1}(\cdot)$. The output of $\mathrm{R}_1$  is a bit sequence:
\begin{eqnarray}
\hspace{-0.5cm}\bm{b}_1 = f_{\mathrm{R}_1}(\bm{y}_1), 
\label{equ:yi}
\end{eqnarray}
{where the length of $\bm{b}_1$, denoted as $L_{\bm{b}_1}$, varies according to different $\bm{y}_1$ implementations which will be further transmitted over the core network.}  

Upon receiving $\bm{b}_1$, the destination reconstructs the original image by a decoding function $f_{\mathrm{D}}(\cdot)$ as $\hat{\bm{S}} = f_{\mathrm{D}}(\bm{b}_1)$. To evaluate the reconstruction performance of different processing strategies (specified by $f_s$, $f_{\mathrm{R}_1}$ and $f_{\mathrm{D}}$), the peak signal-to-noise ratio (PSNR) and the structural similarity index (SSIM) are adopted, the PSNR is defined as:
\begin{equation}
    \text{PSNR} = 10\log_{10} \frac{255^2}{\frac{1}{M}||\bm{S}-\hat{\bm{S}}||^2_F},
    \label{eq:psnr}
\end{equation}
where $M = CHW$ denotes the total number of pixels in the image $\bm{S}$. {The SSIM is defined as:}
\begin{equation}
    \text{SSIM} = \frac{(2\mu_s\mu_{\hat{s}}+c_1)(2\sigma_{s\hat{s}}+c_2)}{(\mu^2_s + \mu^2_{\hat{s}} + c_1)(\sigma^2_s + \sigma^2_{\hat{s}} + c_2)},
    \label{eq:SSIM}
\end{equation}
where $\mu_s, \sigma_s, \sigma_{s\hat{s}}$ are the mean and variance of $\bm{S}$, and the covariance between $\bm{S}$ and $\hat{\bm{S}}$, respectively. $c_1$ and $c_2$ are constants for numeric stability.

\begin{figure*}[t]
\centering
\includegraphics[width=0.85\linewidth]{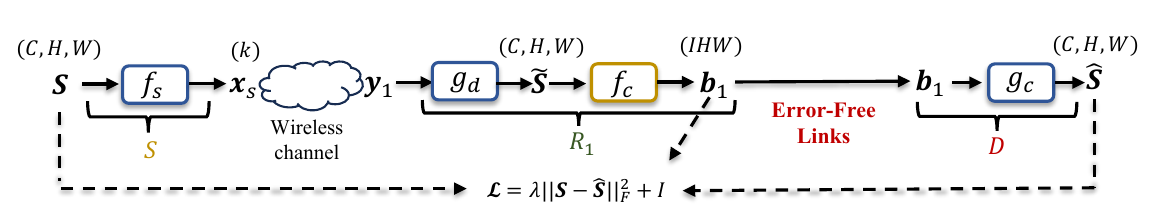}\\
\caption{Illustration of the proposed h-DJSCC framework, where $f_s(\cdot)$ denotes the DeepJSCC encoder at the source node, $g_d(\cdot)$ denotes the DeepJSCC decoder located at $\mathrm{R}_1$. $f_c(\cdot)$ and $g_c(\cdot)$ represent the compression and decompression modules. The bit output, $\bm{b}_1$ of the compression module, $f_c(\cdot)$ is delivered to the destination via error-free links in the core network. We show the dimensions on top of the tensors and vectors for clarity. The loss function, $\mathcal{L}$, is a summation of the weighted distortion and the bpp ($I$) defined in \eqref{equ:Izv} used to compress the image $\widetilde{S}$.}
\label{fig:jsc_framework}
\end{figure*}

\subsection{Existing Methods}
We first review existing methods for communication over multi-hop wireless networks, namely, the fully digital transmission and the naive quantization approach.

\subsubsection{Fully digital transmission}\label{section:fdt} The encoder at $\mathrm{S}$ first compresses the input image into a bit sequence, denoted by $\bm{b}_s$, which is then channel coded and modulated to form the channel codeword $\bm{x}_s$. We consider the state-of-the-art source encoders (e.g., BPG) which output variable-length codewords, thus, the number of complex channel uses $k$ refers to the expected length of $\bm{x}_s$. The processing operation $f_{\mathrm{R}_1}(\cdot)$ at the first relay is simply a channel decoder, denoted as $D_1(\cdot)$, which decodes the received signal, $\bm{y}_1$ to a bit sequence, $\hat{\bm{b}}_s$. The bit sequence will be transmitted over the core network and the destination node reconstructs the image using $\hat{\bm{b}}_s$.

To better illustrate the processing of the fully digital transmission scheme, we provide a concrete example where the number of channel uses, $k = 768$ and $\eta = 2$ dB which corresponds to the channel capacity $\mathcal{C}_s = 1.37$.
We adopt a coded modulation scheme with a rate below $\mathcal{C}_s$ (for reliable transmission). To be precise, we consider a rate-1/2 LDPC code with 4QAM corresponding to an achievable rate $R_s = 1$ for the first hop. It has been shown by simulations that the coded modulation schemes along with the corresponding belief propagation (BP) decoders are capable to achieve almost zero block error rate (BLER) in the considered scenario. {The average number of bits, i.e., $L_{\hat{\bm{b}}_s}$,} to be transmitted over the core network is $768$ and is delivered to the destination losslessly.

A key weakness of the fully digital transmission is that, {the link quality between the source and the first relay node might change over time, and the system suffers from the cliff and leveling effects.}

\subsubsection{Naive quantization}\label{Sec:naive_quant}
{To avoid the cliff and leveling effects, we adopt DeepJSCC for the first hop with analog transmission where $\bm{x}_s$ is a DeepJSCC codeword.} The first relay node transforms its received noisy codeword, $\bm{y}_1$ to a bit sequence, $\bm{b}_1$, and transmits it over the core network. This is analogous to the CPRI compression \cite{scalar_cpri, vq_cpri} in the literature where baseband signals are quantized into bits to be transmitted over the fiber. In this paper, we consider a vector quantization approach. In particular, the received signal $\bm{y}_1$ is first partitioned into $k/N_v$ blocks where $N_v \ge 1$ denotes the number of elements in the block. $N_v b$ number of bits is used to quantize each block and the codebook to quantize each block is obtained via Lloyd algorithm which is available at the first relay, $\mathrm{R}_1$ and the destination. 

Note that the elements within $\bm{y}_1$ are not identically and independently distributed (i.i.d.). Moreover, the underlying distribution is hard to model, thus, the vector quantization scheme is sub-optimal calling for more advanced compression algorithm which will be detailed in the next section.

\section{The proposed h-DJSCC framework}\label{sec:JSC}
\subsection{Hybrid Solution}
We consider a hybrid scheme (see Fig. \ref{fig:jsc_framework}), where the mobile user adopts DeepJSCC protocol in the first hop to avoid the cliff and leveling effects, whereas the first relay node, $\mathrm{R}_1$, uses a DNN to compress its received signal, $\bm{y}_1$ to a bit sequence $\bm{b}_1$ which is transmitted over the $(n-1)$ hops in the core network errorlessly.  Then, the decoder at $\mathrm{D}$, takes $\bm{b}_1$ as input to generate the final reconstruction, $\hat{\bm{S}}$.  {The flowchart of the proposed h-DJSCC framework is detailed as follows.}

\subsubsection{Encoder at $\mathrm{S}$} 
The encoder function $f_s(\cdot)$ at $\mathrm{S}$ is shown in Fig. \ref{fig:fig_NN} which is  comprised of 2d CNN layers with residual connection. 

\subsubsection{Processing at $\mathrm{R}_1$}
A naive solution is to directly quantize the elements of the received signal $\bm{y}_1$ to a bit sequence which is illustrated in Section \ref{Sec:naive_quant}. However, naively quantizing the received signal leads to a sub-optimal solution, which motivates the employment of a neural compression model.  

Instead of directly compressing $\bm{y}_1$, we found it is more beneficial to first transform $\bm{y}_1$ into a tensor with the same dimension of the original image $\bm{S}$, denoted by $\widetilde{\bm{S}}$,  via a DeepJSCC decoder, $g_d(\cdot)$, whose neural network architecture is shown in Fig. \ref{fig:fig_NN}, and then apply the neural image compression algorithm \cite{balle2018variational}, denoted as $f_c(\cdot)$, to compress $\widetilde{\bm{S}}$ into a bit sequence $\bm{b}_1$. The processing of the neural compression model is summarized in Section \ref{sec:pre_img_comp} and is omitted here. Due to the fact that different images contain different amount of information, the length of the bit sequence varies, and, similarly to the fully digital transmission, we evaluate the expected amount of {latency} as 
\begin{equation}
    {k^\prime = 
\frac{1}{R_N} \mathbb{E}_{\widetilde{\bm{S}} \sim p_{\widetilde{\bm{S}}}}(L_{\bm{b}_1}),}
\label{equ:k_prime}
\end{equation}
{where $L_{\bm{b}_1}$ denotes the number of bits in the sequence while $R_N$ is the {achievable rate of the backhaul links.}}

\subsubsection{Neural processing at $\mathrm{D}$} 
{Thanks to the reliable backhaul links, the destination node retrieves the bit sequence $\bm{b}_1$ correctly, and employs arithmetic decoders and the non-linear transformation functions to generate the final reconstruction image $\hat{\bm{S}}$ as illustrated in Section \ref{sec:pre_img_comp}. }

\begin{figure}[!t]
\centering
\includegraphics[width=0.9\columnwidth]{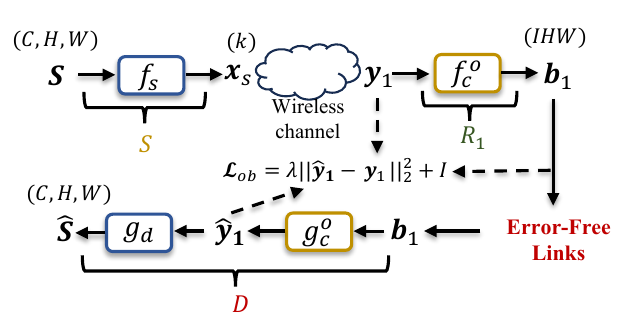}
\caption{The framework of the oblivious relaying where $\mathrm{R}_1$ only has access to the past received signal $\bm{y}_1$.}
\label{fig:fig_oblivious_frame}
\end{figure}

\subsection{Oblivious Relaying}\label{sec:oblivious}
As shown in Fig. \ref{fig:jsc_framework}, the first relay $\mathrm{R}_1$ utilizes the DeepJSCC decoder $g_d(\cdot)$ which is jointly optimized with $f_s(\cdot), f_c(\cdot)$ and $g_c(\cdot)$ to achieve the best reconstruction performance at the destination. This implies that the relay has access to the distribution of the transmitted image and the neural processing function, $f_s(\cdot)$ at the source. However, in some scenarios, the relay is unaware of the exact processing at the source as well as the underlying data distribution\footnote{It might be the case where the source and the destination wish to convey some private message $\bm{S}$ which should not be decoded by the relays.}, $P_{\bm{S}}$, which motivates a different relaying protocol known as oblivious relaying \cite{tit_oblivious, tcom_oblivious} in the literature. 
In particular, the authors in \cite{tit_oblivious} consider  a scenario where the relay maps its received signal to a codeword from a randomly generated codebook which will be forwarded to the destination.

In our case, instead of using a randomly generated codebook, the relay compresses $\bm{y}_1$ using a DNN.  As shown in Fig. \ref{fig:fig_oblivious_frame}, the compressor, $f_{c}^{o}(\cdot)$ is employed at $\mathrm{R}_1$ to compress the received signal into bits and the decompressor, $g_{c}^{o}(\cdot)$, located at the destination, reconstructs the received signal as:
\begin{equation}
    \hat{\bm{y}}_1 = g_c^o(f_c^o(\bm{y}_1)).
    \label{equ:oblivious_y}
\end{equation}
{The compression and decompression modules follow exactly the same with that illustrated in Fig. \ref{fig:fig_hyperprior} where the latent vectors, $\bm{z}$ and $\bm{v}$, are generated and further compressed into bit sequences.}
Since in the considered oblivious relaying setup, the relay only has access to $\bm{y}_1$, the loss function is expressed as:
\begin{equation}
    \mathcal{L}_{ob} = \lambda \|\bm{y}_1 - \hat{\bm{y}}_1\|^2_2 + I,
    \label{eq:loss_fun_obli}
\end{equation}
where $I$ is defined in \eqref{equ:Izv}. Finally, the DeepJSCC decoder, $g_d(\cdot)$ at the destination takes $\hat{\bm{y}}_1$ as input and outputs the reconstructed image $\hat{\bm{S}}$.

\begin{figure}[t]
\centering
\includegraphics[width=\columnwidth]{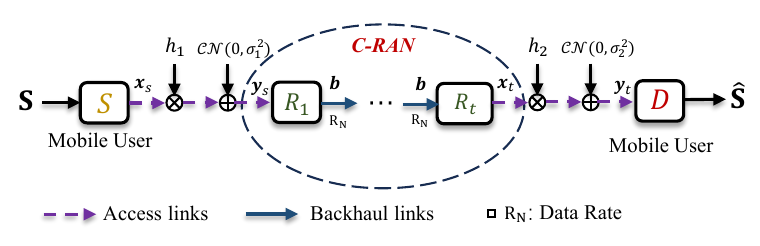}
\caption{{An illustration of the extended h-DJSCC scenario where two mobile users communicate with each other over the wireless network.}}
\label{fig:fig_two_hop_illu}
\end{figure}

{
\subsection{Extension to Two Mobile Users}
As shown in Fig. \ref{fig:fig_two_hop_illu}, we consider a more practical scenario, dubbed `extended h-DJSCC' scenario,  where the destination node is also a mobile user which is connected to the C-RAN via a wireless access link with fluctuating channel qualities.

Since there are only two wireless access links, the noise accumulation problem is less significant even if both hops adopt analog transmission which is verified in \cite{zju_multihop}. Thus, we adopt DeepJSCC to generate the codewords for the two wireless access links. To be precise, the source node encodes the input image, $\bm{S}$, via DeepJSCC encoder, $f_s(\cdot)$, and the first relay node, $\mathrm{R}_1$, decodes the DeepJSCC codeword into a reconstructed image, $\widetilde{\bm{S}}$, using a DeepJSCC decoder, $g_d(\cdot)$.
It then compresses $\widetilde{\bm{S}}$ via a neural compressor, $f_c(\cdot)$, into a bit sequence, $\bm{b}$, which will be transported over the C-RAN to an RRH node, denoted by $\mathrm{R}_t$.
The $\mathrm{R}_t$ first adopts $g_c(\cdot)$ to decompress its received bit sequence, $\bm{b}$, into a reconstructed image, $\bm{\widetilde{S}}_t$:
\begin{equation}
    \widetilde{\bm{S}}_t = g_c(f_c(\widetilde{\bm{S}})),
\end{equation}
and then applies a DeepJSCC encoder, denoted by $f_s^t(\cdot)$, to further encode the image into a DeepJSCC codeword with length $k$ for transmission. The mobile user adopts the DeepJSCC decoder, $g_d^t(\cdot)$, to generate the final reconstructed image, $\bm{\hat{S}}$:
\begin{equation}
    \bm{\hat{S}} = g_d^t(h_2 f_s^t(\widetilde{\bm{S}}_t) + \bm{w}_2),
\end{equation}
where $h_2 \sim \mathcal{CN}(0, 1), \bm{w}_2 \sim \mathcal{CN}(\bm{0}, \bm{I}_k)$ denote the complex fading coefficient and AWGN in the second wireless access link, respectively. 

We parameterize the functions, $f_s(\cdot), g_d(\cdot), f_c(\cdot), g_c(\cdot)$ of the extended h-DJSCC scenario using the weights obtained in Fig. \ref{fig:jsc_framework}. The DeepJSCC encoder and decoder, $f_s^t(\cdot)$ and $g_d^t(\cdot)$, for the second wireless access link can also be parameterized by the weights of $f_s(\cdot), g_d(\cdot)$ in Fig. \ref{fig:jsc_framework}. However, instead of directly adopting the pre-trained models, we find by experiments that fine-tuning $f_s^t(\cdot), g_d^t(\cdot)$ helps to improve the overall performance. Thus, for the extended h-DJSCC scheme, we first load the pretrained neural network weights of the functions, $f_s, g_d, f_c, g_c$, from the h-DJSCC model, then fine-tune $f_s^t, g_d^t$ with $f_s, g_d, f_c, g_c$ fixed\footnote{{It is also possible to fine-tune $f_s, g_d, f_c, g_c$ along with $f_s^t, g_d^t$, yet we find by experiments that this produces identical R-D performance with increased optimization complexity, thus we keep them fixed.}}. The effectiveness of the proposed extended h-DJSCC scheme is evaluated in Section \ref{sec:sim_ext_hdjscc}.
}

\section{Adaptive h-DJSCC Transmission}\label{sec:adaptive_jsc}
In the previous section, for a given $\eta$ value, different models are trained to achieve different points on the R-D trade-off by choosing different $\lambda$ values from a pre-defined set, denoted as $\Lambda$. Since the channel quality between $\mathrm{S}$ and $\mathrm{R}_1$ is {assumed to change over time}, $|\Lambda| |\mathrm{H}|$ different models would be required\footnote{Here we consider the variable $\eta$ is chosen from a discrete set $\mathrm{H}$ with $|\mathrm{H}|$ elements. The proposed fully adaptive h-DJSCC framework can adapt to continuous $\eta$ values.} to achieve satisfactory R-D performance causing severe storage problem. Thus, we would like to develop a fully adaptive h-DJSCC framework that is robust to variations in $\eta$ values and capable of achieving different points on the R-D curve. 

In this section, we introduce the neural network architecture and the corresponding training procedure for the proposed fully adaptive h-DJSCC framework.

\subsection{{R-D Properties of the SNR-adaptive DeepJSCC}}
{As illustrated in Section \ref{sec:pre_sa_module}, the DeepJSCC model with SNR-adaptive (SA) module proposed in \cite{xu2021wireless} achieves satisfactory reconstruction performance by assigning different levels of protection of the information content given the SNR value.}
Despite of its impressive performance, there is no quantitative verification of the concepts such as the information content and the levels of protection which we aim to show it from a R-D perspective.

\begin{figure}[!t]
\centering
\includegraphics[width=0.8\columnwidth]{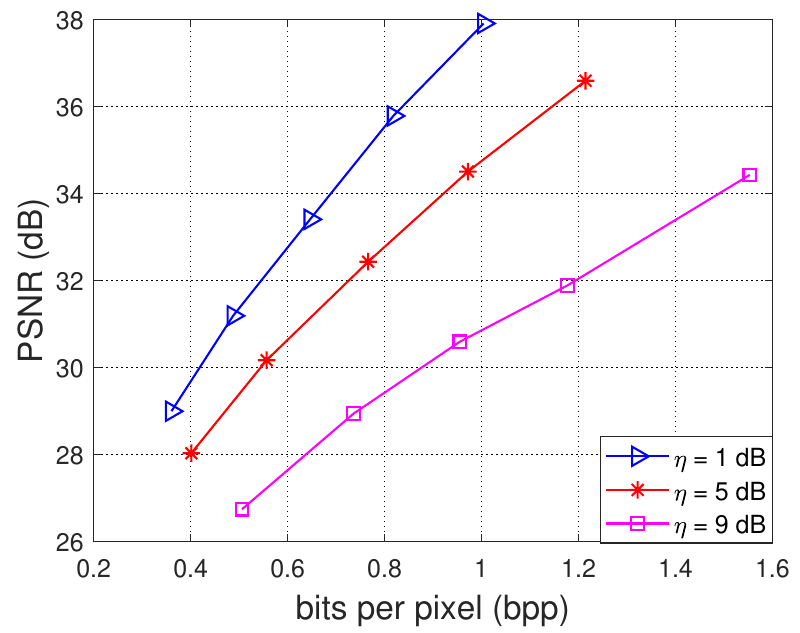}
\caption{{The R-D performance achieved by compressing the output of the SNR-adaptive DeepJSCC decoder, $\bm{S}_{\eta}$, for different $\eta$ values.}}
\label{fig:rdcurve}
\end{figure}

To be specific, we consider an SNR-adaptive DeepJSCC encoder, $f_s(\cdot, \eta)$ which produces different transmitted signals, $\bm{x}_{\eta}$ for different input images and SNR values:
\begin{align}
    \bm{x}_{\eta} = f_s(\bm{S}, \eta).
    \label{equ:snr_adapt_codec}
\end{align}
{The SNR-adaptive DeepJSCC decoder, $g_d(\cdot, \eta)$ produces the reconstructed image as:
\begin{align}
    {\bm{S}}_\eta = g_d(\bm{x}_{\eta}, \eta).
\end{align}}
Then, we compress ${\bm{S}}_\eta$ to bits and decompress the bit sequence to obtain ${\hat{\bm{S}}}_{\eta}$, which is of the same dimension as $\bm{S}_{\eta}$.  Different models corresponding to different $\lambda$ values are trained where the loss function is expressed as:
\begin{equation}
    \mathcal{L}_{SA} = \lambda\|\hat{\bm{S}}_{\eta} - {\bm{S}}_{\eta}\|_2^2 + I,
    \label{equ:loss_fun_curve}
\end{equation}
where $I$ is defined in \eqref{equ:Izv}.
We set $\lambda \in \{200, 400, 800, 1600, 3200\}$ and $\eta \in \{1,5,9\}$ dB for this simulation. As shown in Fig. \ref{fig:rdcurve}, when $\eta$ is high, e.g., $9$ dB, the output of the SNR-adaptive DeepJSCC encoder, $\bm{x}_{\eta}$, contains a larger proportion of information content leading to an inferior R-D curve. On the other hand, when $\eta = 1$ dB, the best R-D curve is obtained showing a large amount of redundancy is produced to protect the information content which aligns with the analysis in \cite{xu2021wireless}.

\begin{figure}[!t]
\centering
\includegraphics[width=0.8\columnwidth]{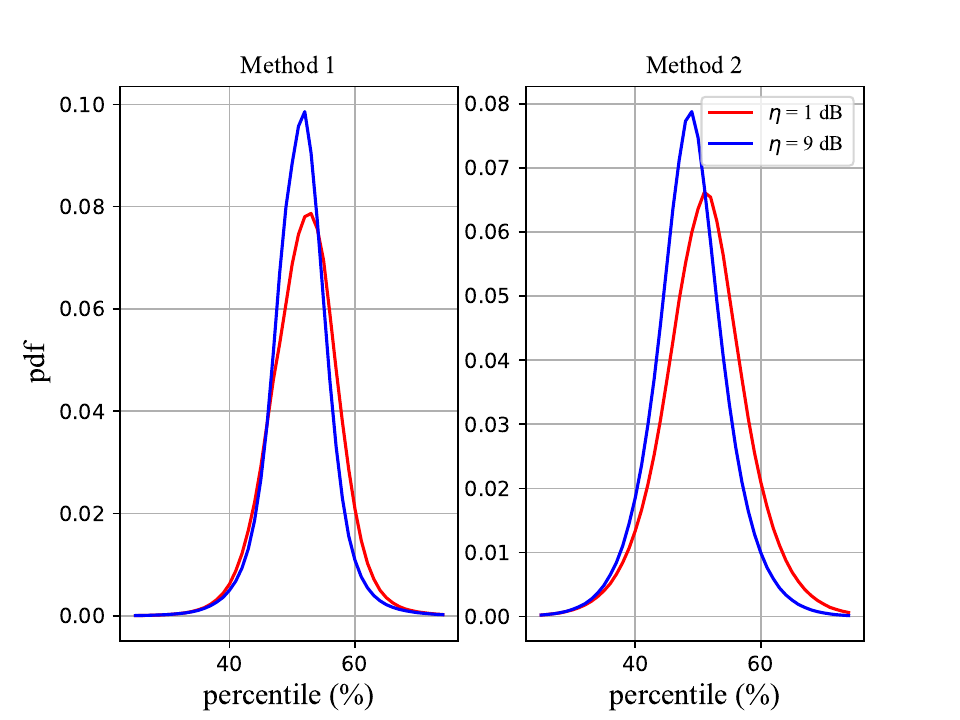}
\caption{The comparison of the empirical distribution, $\hat{P}_{\tilde{\bm{x}}_{\eta}}$, obtained by the h-DJSCC models that is randomly initialized (left) and initialized using a pre-trained SNR-adaptive DeepJSCC model (right). The empirical distributions are evaluated under two different $\eta = \{1, 9\}$ dB values.}
\label{fig:empirical_dist}
\end{figure}

\subsection{SNR-adaptive h-DJSCC Framework}\label{sec:snr_adapt_jsc}
We then focus on the SNR-adaptive neural compression model located at $\mathrm{R}_1$, whose input is the reconstructed image $\widetilde{\bm{S}}_{\eta}$. 
Note that the reconstructed images corresponding to different $\eta$ values follow different distributions. As analyzed in the previous subsection, $\widetilde{\bm{S}}_{\eta}$ corresponding to different $\eta$ values contain different amount of information content. Thus, the SNR-adaptive compression model takes $\eta$ as an additional input compared to the compression model in Section \ref{sec:JSC}. 

We parameterize the analysis transformations, $g_a(\cdot, \eta)$ and $h_a(\cdot, \eta)$, as well as the synthesis transformations, $g_s(\cdot, \eta)$ and $h_s(\cdot, \eta)$ of the SNR-adaptive compression and decompression models by ResNets with SA modules.
The latents $\bm{z}_{\eta}$ and $\bm{v}_{\eta}$ are obtained as:
\begin{equation}
    \bm{z}_\eta = g_a(\widetilde{\bm{S}}_\eta, \eta); \quad \bm{v}_\eta = h_a(\bm{z}_\eta, \eta).
    \label{equ:z_v_eta}
\end{equation}
During training, we focus on estimating the bit budget to encode the above latents. In particular, the pdf of $\tilde{\bm{v}}_{\eta}$ is obtained by replacing $\tilde{\bm{v}}$ in \eqref{equ:probv} by $\tilde{\bm{v}}_{\eta}$:
\begin{align}
p_{\tilde{\bm{{v}}}_{i,\eta}|\bm{\phi}} = \prod_i (p_{\tilde{\bm{{v}}}_{i,\eta}|\phi_i}(\phi_i)*\mathcal{U}(-\frac{1}{2}, \frac{1}{2}))(\tilde{\bm{{v}}}_{i,\eta}),
    \label{equ:prob_snr_v}
\end{align}
where $\tilde{\bm{v}}_{\eta}$ is obtained by adding uniform noise to ${\bm{v}}_{\eta}$.
Note that in the above equation, we make an assumption that the probability model, $p_{\tilde{\bm{{v}}}_{i,\eta}|\phi_i}(\phi_i)*\mathcal{U}(-\frac{1}{2}, \frac{1}{2})$ is capable to provide accurate pdf for the latent variable $\tilde{\bm{v}}_{\eta}$ with varying $\eta$ values.
Similarly, the pdf of $\bm{z}_{\eta}$ can be expressed by replacing  $\tilde{\bm{z}}$ and $\tilde{{\bm{v}}}$ in \eqref{equ:probz} by $\tilde{\bm{z}}_\eta$ and $\tilde{\bm{v}}_\eta$ which is omitted due to the page limit.

It is worth emphasizing that it is also possible to let the probability model $p_{\tilde{\bm{v}}_{\eta}|\bm{\phi}}$ be aware of different SNR values by concatenating the SNR value along with the input tensor, $\tilde{\bm{v}}_{\eta}$. 
{However, we found by experiments that the neural network $h_a(\cdot, \eta)$ is already capable of encoding the varying SNR values into its output, $\tilde{\bm{v}}_{\eta}$ to fit with the subsequent probability prediction model, $p_{\tilde{\bm{{v}}}_{i,\eta}|\bm{\phi}}$ producing accurate bit estimation for different $\eta$ values without explicitly feeding the $\eta$ value to $p_{\tilde{\bm{{v}}}_{i,\eta}|\bm{\phi}}$. Thus, we adopt the formula in \eqref{equ:prob_snr_v} throughout the paper.} 

The SNR-adaptive h-DJSCC framework is optimized in an end-to-end fashion. In particular, for each batch, we sample $\eta \sim \mathcal{U}(\eta_{min}, \eta_{max})$, and the loss corresponding to the sampled SNR value is calculated according to \eqref{eq:loss_fun}.  It is worth emphasizing that the initialization method plays a vital role in achieving satisfactory R-D performances. We found in our experiments that adopting random initialization for all the neural network weights leads to highly sub-optimal solution.

To understand this phenomena, we explore the  distribution of the latent vector, $\bm{x}_{\eta}$. Instead of directly studying the complex vector, we convert it to a real-valued one, denoted by $\bm{x}^\prime_{\eta} \in \mathbb{R}^{2k}$ and normalize its elements as:
\begin{align}
    \tilde{\bm{x}}_{\eta} = \frac{\bm{x}^\prime_{\eta} - \min(\bm{x}^\prime_{\eta})}{\max({\bm{x}^\prime_{\eta}) - \min(\bm{x}^\prime_{\eta})}}.
    \label{equ:prob_norm}
\end{align}
To estimate the empirical distribution, $\hat{P}_{\tilde{\bm{x}}_{\eta}}(\cdot)$, we partition the unit interval to $100$ bins, and calculate the number of elements that fall into each bin. $N = 10^4$  realizations of $\tilde{\bm{x}}_{\eta}$ are collected and $\hat{P}_{\tilde{\bm{x}}_{\eta}}(\cdot)$ is obtained by averaging over all $2kN$ elements:
\begin{align}
    \hat{P}_{\tilde{\bm{x}}_{\eta}}(i) = \frac{1}{2kN} \sum_{n=1}^{N}\sum_{j=1}^{2k} \bm{1}\left(\frac{i-1}{100} < {\tilde{\bm{X}}_{\eta}[n,j]}\le \frac{i}{100}\right),
    \label{equ:empirical}
\end{align}
where $\tilde{\bm{X}}_{\eta} \in \mathbb{R}^{N\times 2k}$ is a matrix obtained by stacking $N$ different realizations and $\bm{1}(\cdot)$ is the indicator function.

On the left hand side of Fig. \ref{fig:empirical_dist}, the empirical distribution $\hat{P}_{\tilde{\bm{x}}_{\eta}}$ for the SNR-adaptive h-DJSCC framework with random neural network weight initialization is plotted for $\eta = \{1, 9\}$ dB. It can be seen that the two distributions are quite close to each other. This implies that the scheme converges to a sub-optimal solution as the two distributions need to be more distinct w.r.t different channel qualities.

{
However, it is hard to explicitly design a loss function to make the output distributions corresponding to different SNR values more distinguishable. Adopting the Kullback–Leibler (KL) divergence defined as:
\begin{equation}
    D_{\mathrm{KL}}(P \| Q) = \sum_{i} P(i) \log \frac{P(i)}{Q(i)},
\end{equation}
is a possible solution where $P, Q$ denote two probability distributions. However, it requires updating the probability distributions corresponding to different $\eta$ values for each and every optimization step leading to prohibitive training complexity. 
Fortunately, we find by experiments that the SNR-adaptive DeepJSCC encoder (without being jointly trained with the compression models) produces the desired output distribution. Thus, we initialize the DeepJSCC modules of the SNR-adaptive h-DJSCC model using the pre-trained SNR-adaptive DeepJSCC model. As will be shown in the simulation section, the above training approach fulfills the SNR-adaptive objective. 
Moreover, the KL divergence of the two empirical distributions, $P_{\tilde{\bm{x}}_{1}}$ and $P_{\tilde{\bm{x}}_{9}}$ obtained from the random initialization and a pre-trained model are $0.04$ and $0.09$, respectively. This implies that the pre-trained model yields more `distinct' output distribution w.r.t. different SNR values which can also be verified visually from Fig. \ref{fig:empirical_dist}.
}

\begin{figure*}[!t]
\centering
\includegraphics[width=0.9\linewidth]{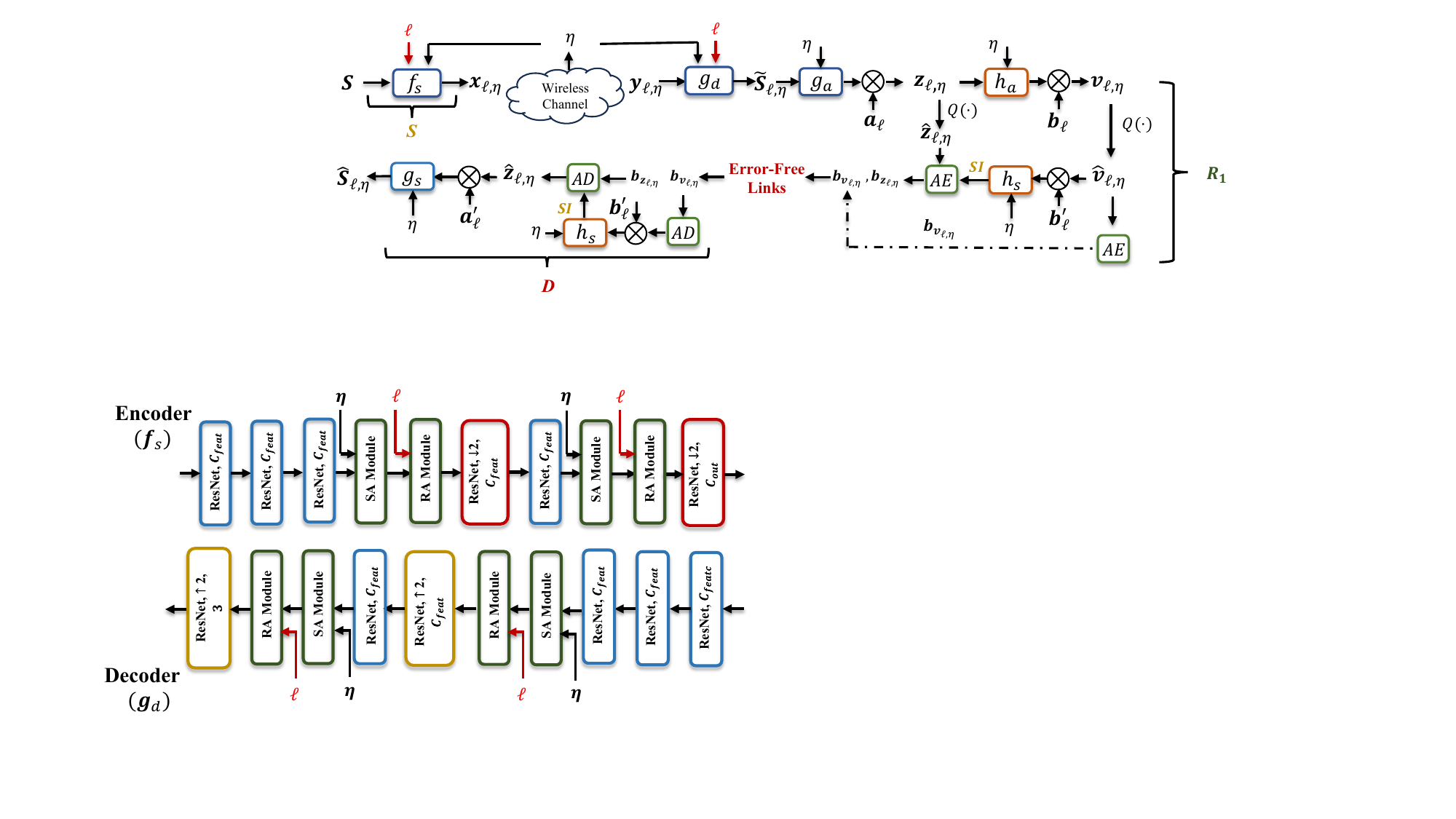}\\
\caption{{The flowchart of the proposed fully adaptive h-DJSCC framework. It is applicable to both AWGN and fading channels where SA and RA modules are adopted to achieve the SNR and rate-adaptive objectives, respectively. The  $AE$ and $AD$ denote the arithmetic encoding and decoding blocks, respectively. We consider the R-D performance corresponding to $\lambda_\ell$ with the scaling factors $\{\bm{a}_\ell, \bm{a}^\prime_\ell, \bm{b}_\ell, \bm{b}^\prime_\ell\}$. Finally,  $\text{SI}$ is used to represent $\bm{\mu}_{\ell, \eta}$ and $\bm{\sigma}^2_{\ell, \eta}$.}}
\label{fig:flowchart}
\end{figure*}

\IEEEoverridecommandlockouts

\newcommand{\y}{\bm{y}}

\begin{algorithm}[!h]
	\caption{Training procedure of the proposed fully adaptive h-DJSCC framework over a fading channel.}
    \label{alg:full_adapt_jsc}

	\SetKwInOut{Input}{Input}\SetKwInOut{Output}{Output}

	\Input{$\{\bm{S}\}_{1:N}, \Lambda, \eta_{min},                        \eta_{max}, Epochs$}
	\Output{$\hat{\bm{S}}_{\ell, \eta}, I_{\ell, \eta}$}
	
	\BlankLine

	\textbf{Initialization:}\\{
{Initialize $f_s(\cdot, \eta, \ell), g_d(\cdot, \eta, \ell)$ using pre-trained model \cite{xu2021wireless}.}\\
	 Randomly initialize $\{f_c(\cdot, \eta), g_c(\cdot, \eta)\}$ consist of $
  \{\bm{A}, \bm{A}^\prime, \bm{B}, \bm{B}^\prime, g_a(\cdot, \eta), g_s(\cdot, \eta), h_a(\cdot, \eta), h_s(\cdot, \eta)\}$\\
         }

        \%\% \textbf{Training Phase}\\
        \For{$t=1$ \KwTo {$Epochs$}}{
	\For{$b=1$ \KwTo {$\mathcal{B}$}}{
	    {Sample $\lambda_\ell \sim \text{Uniform}(\Lambda), \eta \sim \mathcal{U}(\eta_{min}, \eta_{max})$}, $h \sim \mathcal{CN}(0, 1)$\\
	\% \textbf{SNR-adaptive DeepJSCC model:}\\

        \eIf{with CSIT}{
    {$\bm{x}_{\ell, \eta} = \frac{h^*}{|h|} f_s(\bm{S}, |h|^2\eta, \ell)$}. \Comment{Source Node}\\
    {$\bm{y}_{\ell, \eta} = h\bm{x}_{\ell, \eta} + \bm{w}$},\Comment{Channel} \\
    {$\hat{\bm{x}}_{\ell, \eta} = \frac{|h| \bm{y}_{\ell, \eta}}{|h|^2 + 1/\eta}$.}\\
    }
    {{$\bm{x}_{\ell, \eta} = f_s(\bm{S}, \eta, \ell)$}.  \\
    {$\bm{y}_{\ell, \eta} = h\bm{x}_{\ell, \eta} + \bm{w}$}, \\
    {$\hat{\bm{x}}_{\ell, \eta} = \frac{h^* \bm{y}_{\ell, \eta}}{|h|^2 + 1/\eta}$.}\\
    }

            ${\bm{\widetilde{S}}_{\ell, \eta} = g_d(\hat{\bm{x}}_{\ell, \eta}, \eta, \ell).}$ \Comment{First Relay}\\

            \% \textbf{SNR-adaptive and variable rate compression:} \\
            $\bm{z}_{\ell, \eta} = {g_a(\bm{\widetilde{S}}_{\ell, \eta}, \eta)} \otimes \bm{a}_\ell$, \\
            $\bm{v}_{\ell, \eta} = h_a(\bm{z}_{\ell, \eta}, \eta) \otimes \bm{b}_\ell$, \\
            ${\tilde{\bm{z}}}_{\ell, \eta}, {\tilde{\bm{v}}}_{\ell, \eta} = \bm{z}_{\ell, \eta} + \mathcal{U}(-\frac{1}{2}, \frac{1}{2}), \bm{v}_{\ell, \eta} + \mathcal{U}(-\frac{1}{2}, \frac{1}{2})$,\\
            $\tilde{\bm{z}}^\prime_{\ell, \eta} = \tilde{\bm{z}}_{\ell, \eta} \otimes \bm{a}^\prime_\ell$, \\
            $\hat{\bm{S}}_{\ell, \eta} = g_s(\bm{\tilde{z}}^\prime_{\ell, \eta}, \eta)$,\\
             \textbf{Estimating entropy}:\\
             $\tilde{\bm{v}}^\prime_{\ell, \eta} = \tilde{\bm{v}}_{\ell, \eta} \otimes \bm{b}^\prime_\ell$, \\
             $\bm{\tilde{\mu}}_{\ell, \eta}, \bm{\tilde{\sigma}}_{\ell, \eta} = h_s(\bm{\tilde{v}}^\prime_{\ell, \eta}, \eta)$,\\
             $I_{v, \ell, \eta} = -\log_2(p_{\tilde{\bm{v}}_{\ell, \eta}})$, \\
             $I_{z, \ell, \eta} = -\log_2(p_{\tilde{\bm{z}}_{\ell, \eta}})$,
             \Comment{$p_{\tilde{\bm{v}}_{\ell, \eta}}$ and $p_{\tilde{\bm{z}}_{\ell, \eta}}$ are calculated in \eqref{equ:prob_snr_v_l} using $\bm{\tilde{\mu}}_{\ell, \eta}$ and $\bm{\tilde{\sigma}}_{\ell, \eta}$.}\\
             $I_{\ell, \eta} = I_{z, \ell, \eta} + I_{v, \ell, \eta}$,\\
            
            \% \textbf{Loss Function:}\\
            $\mathcal{L}_{fa} =   \lambda_\ell \|\bm{S} - \hat{\bm{S}}_{\ell, \eta}\|_F^2 + I_{\ell, \eta}$,\\
            {Optimize parameters in $f_c(\cdot, \eta), g_c(\cdot, \eta), f_s(\cdot, \eta, \ell), g_d(\cdot, \eta, \ell)$ via gradient descent.}
	}}
        
\end{algorithm}

\subsection{{Fully Adaptive h-DJSCC Framework}}\label{sec:full_adapt_jsc}
Finally, we introduce the fully adaptive h-DJSCC framework. We will first discuss the variable rate h-DJSCC framework with a fixed $\eta$ value, inspired by \cite{qsfactor, cui2021asymmetric, modulated_ae} for image compression.
In particular, \cite{qsfactor, cui2021asymmetric, modulated_ae} mitigate the problem by introducing extra learnable parameters called scaling factors for each point on the R-D curve. The scaling factors for each R-D point scale the latent tensors in a channel-wise manner following the intuition that, different channels of the latent tensors are of different levels of importance, i.e., some channels may contain low-frequency components of the image while the others may be comprised of high-frequency component corresponding to the fine-grained features. When a lower rate is required, the channels containing the low-frequency features should be emphasized. When a higher rate is allowed, we can allocate more bits to represent high-frequency features.

Following the procedure described in \cite{cui2021asymmetric}, we introduce four sets of weights, namely, $\{\bm{A}, \bm{A}^\prime, \bm{B}, \bm{B}^\prime\} \triangleq \{\bm{a}_\ell, \bm{a}^\prime_\ell, \bm{b}_\ell, \bm{b}^\prime_\ell,\}, \ell \in [1, |\Lambda|]$. In particular, $\bm{a}_\ell = \{a_{\ell,1}, \ldots, a_{\ell, C_z}\}$ and $\bm{a}^\prime_\ell = \{a^\prime_{\ell,1}, \ldots, a^\prime_{\ell,C_z}\}$ are employed for the output of the $g_a(\cdot)$ module, $\bm{z}$, and the input to the $g_s(\cdot)$ module, $\hat{\bm{z}}_\ell$, respectively\footnote{We first ignore the $\eta$ subscript of the latent vectors $\bm{z}$ and $\bm{v}$ for simplicity.}.  In particular, the scaling and rescaling operations can be expressed as:
\begin{align}
    \bm{z}_{\ell} &= \bm{z}\otimes\bm{a}_\ell, \notag\\
    \bm{\hat{z}}^\prime_\ell &= \bm{\hat{z}}_\ell \otimes \bm{a}^\prime_\ell,
    \label{equ:weight_main}
\end{align}
where $\otimes$ denotes channel-wise multiplication, $\bm{\hat{z}}_\ell$ denotes the quantized version of $\bm{z}_\ell$ and $\bm{\hat{z}}^\prime_\ell$ is the rescaled tensor which is fed to $g_s(\cdot)$ to generate the reconstructed image. 
The hyper latents, $\bm{v}$ and $\hat{\bm{v}}_\ell$ also experience the same scaling and rescaling processes using $\bm{b}_\ell = \{b_{\ell,1}, \ldots, b_{\ell,C_v}\}$ and $\bm{b}^\prime_\ell = \{b^\prime_{\ell,1}, \ldots, b^\prime_{\ell,C_v}\}$. 
The training of the variable rate h-DJSCC framework follows similar procedure as in Section \ref{sec:JSC}. We further note that each set of the weights, $ \{\bm{a}_\ell, \bm{a}^\prime_\ell, \bm{b}_\ell, \bm{b}^\prime_\ell\}$ are optimized for a specific $\lambda_\ell$. 

To gain more insights into the scaling operations, we point out that for a large bit budget (corresponding to a large $\lambda_\ell$ value), the elements in $\bm{a}_\ell$ will generally be larger. This can be understood from the quantization process, when a small $a_{\ell, i}$ value is multiplied with the $i$-th channel, $\bm{z}_i$, the elements of the quantized output, $\bm{\hat{z}}_{\ell, i}$ can only be chosen from a small set of integers resulting in less bits to compress. When a large $a_{\ell, i}$ value is adopted, the candidate set of $\bm{\hat{z}}_{\ell, i}$ values are larger, leading to a better reconstruction performance. We further note that due to the difference among the channels, in general, the relation that $\bm{a}_{\ell_1} = c \bm{a}_{\ell_2}$ does not hold for $\ell_1 \neq \ell_2$, where $c$ is a constant; that is, the weights do not grow linearly with the bit budget.

Finally, we introduce the fully adaptive model where the $\lambda$ values are chosen from a pre-defined candidate set $\Lambda$ and $\eta$ varies within the range $[\eta_{min}, \eta_{max}]$.
To prevent the fully adaptive h-DJSCC framework from converging to a sub-optimal solution as discussed in Section \ref{sec:snr_adapt_jsc}, we first train the SNR-adaptive DeepJSCC model with a varying $\eta \in [\eta_{min}, \eta_{max}]$ and use it to initialize the DeepJSCC encoder and decoder of the fully adaptive h-DJSCC framework.
The weights of the compression and decompression modules of the fully adaptive h-DJSCC framework are initialized randomly. 

{It is worth noticing the influence of the variable rate image compression model on the DeepJSCC model. In principle, a better R-D performance can be obtained if the rate information of the backhaul links, $R_N$, is given to the DeepJSCC encoder and decoder. When $R_N$ is small, the DeepJSCC encoder can assign larger transmit power to the more important features of the image to combat with the channel noise. However, since we are not able to control the exact bpp ($R_N$) achieved by the compression module, we instead feed the index $\ell$ of the $\lambda_\ell \in \{\Lambda\}$ values to the DeepJSCC encoder/decoder. Following this intuition, a RA module is proposed for the DeepJSCC encoder/decoder in the access link whose processing is similar with the SA module described in \eqref{equ:snr_adapt} and their only difference is that the RA module takes the rate information, $\ell$, as input instead of the SNR value. The neural network architecture of the DeepJSCC encoder/decoder for the fully adaptive h-DJSCC model is shown in Fig. \ref{fig:fig_NN} where the proposed RA modules are placed after the SA modules. The DeepJSCC and the image compression models are optimized jointly in an end-to-end fashion.}

The flowchart of the fully adaptive h-DJSCC framework is shown in Fig. \ref{fig:flowchart}. {To start with, the DeepJSCC encoder at the source node $\mathrm{S}$ takes the image $\bm{S}$, the SNR, $\eta$, and the rate information, $\ell$, as input to generate the transmitted signal, $\bm{x}_{\ell, \eta}$. After passing the channel, the DeepJSCC decoder at $\mathrm{R}_1$ takes the received signal, $\bm{y}_{\ell, \eta}$, the SNR value and the rate information to produce the reconstructed image, $\widetilde{\bm{S}}_{\ell, \eta}$, which will be further processed by the SNR-adaptive variable rate compression module, $f_c(\cdot, \eta)$ to generate the latent tensors. }
We multiply the corresponding weights, $\bm{a}_\ell$ and $\bm{b}_\ell$, with the latent tensors in a channel-wise manner to obtain ${\bm{z}}_{\ell, \eta}$ and ${\bm{v}}_{\ell, \eta}$. The scaled tensors will be quantized and arithmetically encoded into a bit sequence, which is then transmitted over the error free links\footnote{The noise over the hops is assumed to be removed via channel codes.}. At the destination node, $\mathrm{D}$, $\hat{\bm{v}}_{\ell, \eta}$ is first arithmetically decoded and rescaled using the weight, $\bm{b}^\prime_\ell$.
Then, $\hat{\bm{z}}_{\ell, \eta}$ is arithmetically decoded with ${\bm{\mu}}_{\ell, \eta}, {\bm{\sigma}}_{\ell, \eta}$ as additional input which are generated using $\hat{\bm{v}}_{\ell, \eta}$. 
Finally, we rescale $\hat{\bm{z}}_{\ell, \eta}$ using $\bm{a}^\prime_\ell$ and the decompression module, $g_c(\cdot, \eta)$ takes the rescaled $\hat{\bm{z}}^\prime_{\ell, \eta}$ as input to generate the reconstructed image, $\hat{\bm{S}}_{\ell, \eta}$. The overall loss function for the fully adaptive h-DJSCC framework can be expressed as:
\begin{align}
    \mathcal{L}_{fa} = \mathbb{E}_{\lambda_\ell \sim \Lambda, \, \eta \sim \mathcal{U}(\eta_{min},  \eta_{max})} \lambda_\ell \|\bm{S} - \hat{\bm{S}}_{\ell, \eta}\|_F^2 + I_{\ell, \eta},
    \label{equ:full_loss_fun}
\end{align}
where $I_{\ell, \eta}$ can be estimated as follows:
\begin{align}
    &I_{\ell, \eta} = \frac{1}{HW} \mathbb{E}_{\tilde{\bm{z}}_{\ell, \eta}, \tilde{\bm{v}}_{\ell, \eta}\sim q}\notag \\ &\left[-\log_2(p_{\tilde{\bm{z}}_{\ell, \eta}|\tilde{\bm{v}}_{\ell, \eta}}(\tilde{\bm{z}}_{\ell, \eta}|\tilde{\bm{v}}_{\ell, \eta}))  - \log_2(p_{\tilde{\bm{v}}_{\ell, \eta}}(\tilde{\bm{v}}_{\ell, \eta})) \right].
\label{equ:I_lsnr}
\end{align}
Note that the $p$ terms denote the prior distributions that are optimized during training while $q$ terms represent the posterior distribution given $\tilde{\bm{S}}_{\eta}$. We provide the exact forms of the $p$ terms which resemble that in \eqref{equ:probz} and \eqref{equ:probv}:
\begin{align}
p(\tilde{\bm{z}}_{\ell, \eta}|\tilde{\bm{v}}_{\ell, \eta}) &= \prod_i (\mathcal{N}(\tilde{{\mu}}_{{\ell, \eta}, i}, \tilde{\sigma}^2_{{\ell, \eta}, i})*\mathcal{U}(-\frac{1}{2}, \frac{1}{2}))(\tilde{\bm{z}}_{{\ell, \eta}, i}) \notag \\
&\widetilde{\bm{\mu}}_{\ell, \eta}, \widetilde{\bm{\sigma}}_{\ell, \eta} = h_s(\tilde{\bm{v}}_{\ell, \eta},  \eta, \bm{\theta}) \notag \\
p_{\tilde{\bm{v}}_{\ell, \eta}|\bm{\phi}} &= \prod_i (p_{\tilde{\bm{v}}_{\ell, \eta, i}|\phi_i}*\mathcal{U}(-\frac{1}{2}, \frac{1}{2}))(\tilde{\bm{v}}_{\ell, \eta, i}).
    \label{equ:prob_snr_v_l}
\end{align}
as well as the $q$ term which is similar with that in \eqref{equ:post}:
\begin{align}
    q(\tilde{\bm{z}}_{\ell, \eta}, \tilde{\bm{v}}_{\ell, \eta}|\bm{\widetilde{S}}_{\eta}) &= \notag  \prod_i \mathcal{U}(\tilde{\bm{z}}_{\ell, \eta, i}|\bm{z}_{\ell, \eta, i}-\frac{1}{2}, \bm{z}_{\ell, \eta, i}+\frac{1}{2}) \\ &\mathcal{U}(\tilde{\bm{v}}_{\ell, \eta, i}|\bm{v}_{\ell, \eta, i}-\frac{1}{2}, \bm{v}_{\ell, \eta, i}+\frac{1}{2}),
\label{equ:post_l}
\end{align}
where $\bm{z}_{\ell, \eta, i}$ denotes the $i$-th channel of $\bm{z}_{\ell, \eta}$.

\subsection{{Fading Channel}}\label{sec:fading}
Finally, we evaluate the fully adaptive h-DJSCC framework over fading channels. To be specific, the link between $\mathrm{S}$ and $\mathrm{R}_1$ is replaced by a Rayleigh fading channel while the remaining hops within the core network are kept the same. {We first consider the case with fixed $\eta$ and $\ell$ values.}

Two different setups of the Rayleigh fading channel are considered, in the first case, we assume the channel state information (CSI), denoted by $h$, is available to both $\mathrm{S}$ and $\mathrm{R}_1$. In the second case, however, only the receiver has access to the CSI. 

{In the first scenario, the source node utilizes the CSI and the rate index, $\ell$, as additional inputs to generate its channel input vector, $\bm{x}_s$:}
\begin{equation}
    {{\bm{x}}_s = f_s(\bm{S}, |h|^2 \eta, \ell),}
    \label{equ:csit_xs}
\end{equation}
where $|h|^2 \eta$ is the effective SNR. It then precodes ${\bm{x}}_s$ as:\newline \newline \newline
\begin{equation}
    \tilde{\bm{x}}_s = \frac{h^*}{|h|} \bm{x}_s.
    \label{equ:precode}
\end{equation}
The received signal at $\mathrm{R}_1$ can be expressed as:
\begin{equation}
    \bm{y}_1 = h \tilde{\bm{x}}_s + \bm{w}_s =  {|h|} \bm{x}_s + \bm{w}_s,
    \label{equ:fading}
\end{equation}
where the fading coefficient, $h$ follows a complex Gaussian distribution, i.e., $h \sim \mathcal{CN}(0, 1)$ and $\bm{w}_s \sim \mathcal{CN}(\bm{0}, \frac{1}{\eta} \bm{I}_k)$. The first relay node receives $\bm{y}_1$ and applies the minimum mean square error (MMSE) equalization to estimate $\bm{x}_s$:
\begin{equation}
    \hat{\bm{x}}_s =  \frac{|h| \bm{y}_1}{|h|^2 + 1/\eta}.
    \label{equ:mmse}
\end{equation}

For the second scenario, {the source generates $\bm{x}_s = f_s(\bm{S}, \eta, \ell)$} and directly transmits it to the channel as it has no access to CSI. The MMSE equalization in this case is expressed as: 
\begin{equation}
    \hat{\bm{x}}_s =  \frac{h^* \bm{y}_1}{|h|^2 + 1/\eta}.
    \label{equ:mmse2}
\end{equation}
For both setups, the output, $\hat{\bm{x}}_s$ is fed to the subsequent DeepJSCC decoder and the compression modules which are identical to the AWGN case. Moreover, we note that the fully adaptive scheme depicted in Section \ref{sec:full_adapt_jsc} for AWGN channel is also applicable to the fading scenario. The overall training procedure of the fully adaptive h-DJSCC framework for the Rayleigh fading channel is summarized in Algorithm \ref{alg:full_adapt_jsc}.


\section{Numerical Experiments}

\label{sec:experiment}
\subsection{Parameter Settings and Training Details}
Unless otherwise mentioned, the number of complex channel uses for the first hop is fixed to $k = 768$, which corresponds to $C_{out}=24$. The number of channels, $C_z$ and $C_v$, for $\bm{z}$ and $\bm{v}$ are set to 256 and 192, respectively.

Similar to the training setting in \cite{bian2024processandforward}, we consider the transmission of images from the CIFAR-10 dataset, which consists of $50,000$ training and $10,000$ test images with $32 \times 32$ resolution. Adam optimizer is adopted with a varying learning rate, initialized to $10^{-4}$ and dropped by a factor of $0.8$ if the validation loss does not improve in $10$ consecutive epochs.

\subsection{Performance Evaluation}

\begin{figure}[!t]
\centering
\includegraphics[width=0.8\columnwidth]{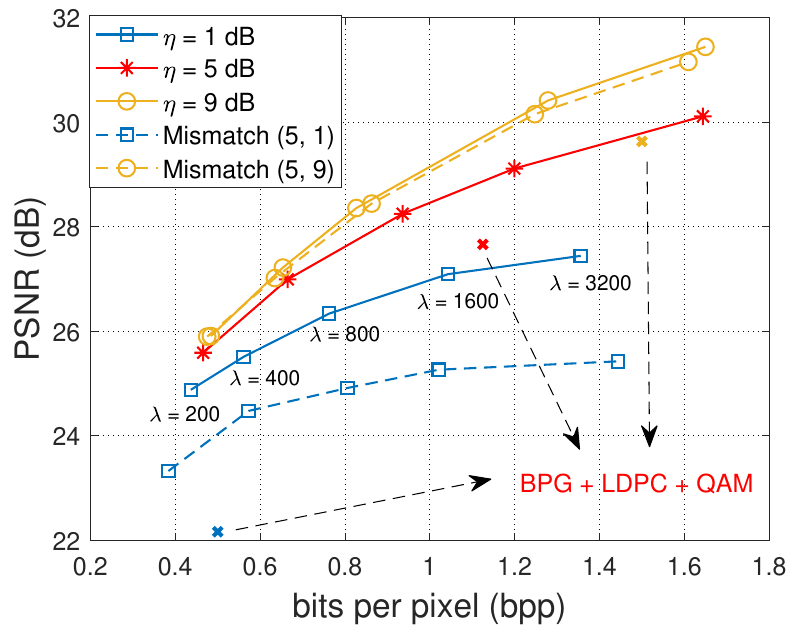}
\caption{Solid lines: R-D tradeoff for the proposed h-DJSCC framework. 15 different h-DJSCC models are trained and evaluated corresponding to $\eta \in \{1, 5, 9\}$ dB and $\lambda \in \{200, 400, 800, 1600, 3200\}$. Dashed lines: Models trained at $\eta = 5$ dB are evaluated under $\eta = \{1, 9\}$ dB.}
\label{fig:fig_rdtradeoff_w_cliff}
\end{figure}

\subsubsection{R-D performance for the proposed h-DJSCC framework} 
Next, we focus on the proposed h-DJSCC framework and show that by training models with different $\lambda$ values, the h-DJSCC framework can achieve different points on the R-D plane, enabling flexible adaptation to the user specific reconstruction requirement and latency constraint.  Different $\eta \in \{1, 5, 9\}$ dB  are evaluated and for each of the SNR value, we train different models with respect to different $\lambda \in \{200, 400, 800, 1600, 3200\}$ values. 

The solid curves in Fig. \ref{fig:fig_rdtradeoff_w_cliff} represent the PSNR performance versus bpp $I$ defined in \eqref{equ:Izv}. For all $\eta$ values, different bpp values are obtained with different models. When the latency constraint of the destination is stringent, we adopt a model with small $\lambda$, when the destination requires a better reconstruction performance, however, a large $\lambda$ is preferable. We also find a better R-D curve is obtained when $\eta$ is larger, which is intuitive as the first hop becomes less noisy. 

The R-D performance of the fully digital and the naive quantization scheme is also evaluated. In particular, for the fully digital scheme, we consider BPG compression algorithm followed by LDPC code and QAM modulation with different code rates (e.g., $1/2, 3/4, 5/6$) and modulation orders (e.g., BPSK, 4QAM, 16QAM, 64QAM) for reliable transmission. For each $\eta$ value, a specific code rate and modulation order which achieves the best R-D performance is adopted whose performance is shown in Fig. \ref{fig:fig_rdtradeoff_w_cliff}. As can be seen, for all $\eta$ values, the proposed h-DJSCC framework achieves better R-D performance compared with the fully digital scheme. The performance of the naive quantization scheme is evaluated under $\eta = 5$ dB with $N_v = 2, b = 1$ whose PSNR equals to $18.8$ dB and bpp equals to $1.5$. Though naive quantization scheme is capable to avoid the cliff and leveling effects, its PSNR performance is outperformed by both the proposed h-DJSCC and the fully-digital baseline. 

\subsubsection{Avoiding the cliff and leveling effects} 
We then show that the proposed h-DJSCC framework is capable to avoid the cliff and leveling effects.
In this simulation, models corresponding to different $\lambda$ values trained at $\eta = 5$ dB are evaluated at $\eta = \{1, 9\}$ dB whose performances are represented by the dashed lines in Fig. \ref{fig:fig_rdtradeoff_w_cliff}.  As shown in the figure, reasonable performance is maintained with mismatched SNR values which verifies the effectiveness of the proposed h-DJSCC framework.

\begin{figure}[!t]
\centering
\includegraphics[width=0.8\columnwidth]{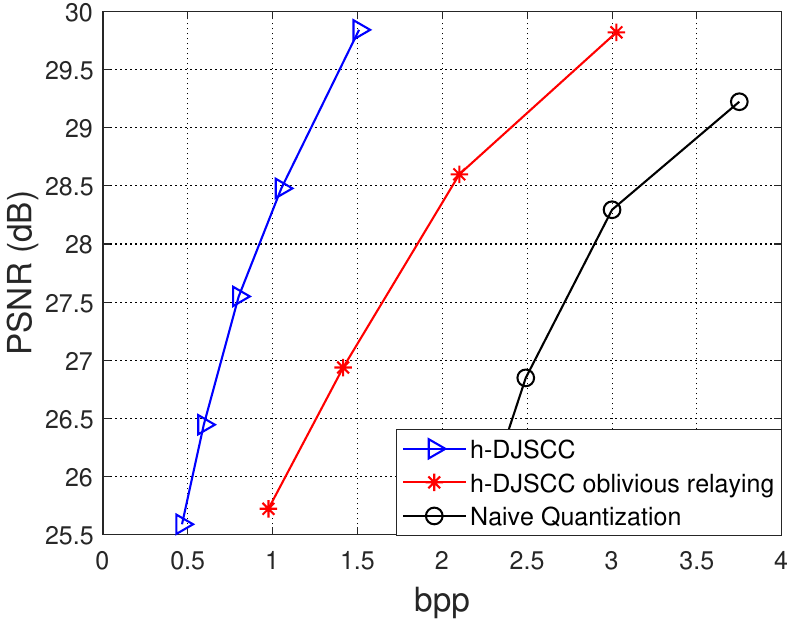}
\caption{Comparison of the R-D performance achieved by the h-DJSCC framework with/without oblivious relaying and the naive quantization scheme. In this simulation, we set $k = 768$, $\eta = 5$ dB.}
\label{fig:fig_oblivious}
\end{figure}

\subsubsection{Oblivious relaying}\label{sec:sim_oblivious} 
We provide simulation results of the oblivious relaying mentioned in Section \ref{sec:oblivious} where the relay $\mathrm{R}_1$ only has access to the past received signal $\bm{y}_1$. We set $k = 768$, $\eta = 5$ dB and $\lambda = \{5, 7.5, 15, 25\}$ for this simulation. The relative performance of the oblivious relaying with the original h-DJSCC framework is shown in Fig. \ref{fig:fig_oblivious}.

As expected, the R-D curve obtained from the oblivious relaying is worse than that of the original h-DJSCC framework. {This can be understood from a goal-oriented compression perspective \cite{Deniz2022, kalfa2021towards}} as the neural compression and decompression modules, $f_c(\cdot)$ and $g_c(\cdot)$ of the h-DJSCC framework are jointly optimized for the final image reconstruction, whereas $f_c^o(\cdot)$ and $g_c^o(\cdot)$ of the oblivious relaying model are trained to minimize the reconstruction error of the received signal $\bm{y}_1$. Finally, we show the effectiveness of the proposed h-DJSCC framework under oblivious relaying protocol by comparing it with vector quantization baselines with $N_v = 2, b = \{1.5, 2, 2.5\}$ and $N_v = 3, b = 5/3 $. As can be seen, the h-DJSCC scheme outperforms the naive quantization baseline by a bpp value greater than one.

{ 
\subsubsection{Extended h-DJSCC scheme} \label{sec:sim_ext_hdjscc}

\begin{figure}[!t]
\centering
\includegraphics[width=0.8\columnwidth]{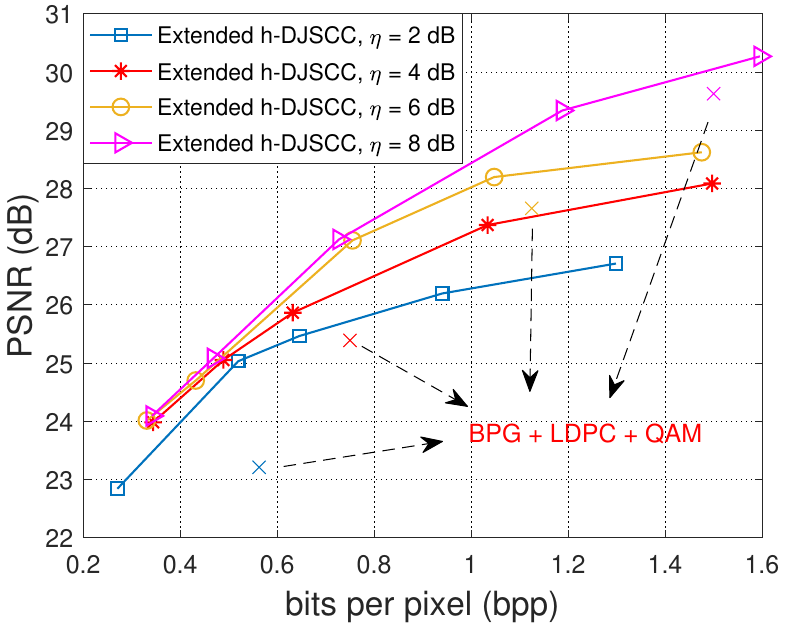}
\caption{{The R-D performance of the extended h-DJSCC scheme with two wireless access links. The BPG compression algorithm followed by different coded modulation schemes are provided as benchmarks.}}
\label{fig:fig_two_hop_sim}
\end{figure}

Finally, numerical experiments are performed to evaluate the effectiveness of the extended h-DJSCC scheme. In this simulation, we consider a simplified scenario where the two wireless access links are assumed to be AWGN channel with identical channel condition, i.e., $h_1 = h_2 = 1$ and $\eta_1 = \eta_2 \triangleq \eta \in \{2, 4, 6, 8\}$ dB\footnote{{We assume $\eta_1 = \eta_2$ for simplicity, however, the proposed extended h-DJSCC scheme is also effective to the $\eta_1 \neq \eta_2$ cases.}}. For each $\eta$ value, we train five different extended h-DJSCC models with $\lambda \in \{200, 400, 800, 1600, 3200\}$. The digital baseline transmits the BPG compression output using different coded modulations. Their relative R-D performances are shown in Fig. \ref{fig:fig_two_hop_sim}.
As can be seen, for each $\eta$ value, the proposed extended h-DJSCC scheme outperforms the digital baseline showing the robustness of the proposed scheme.
}

\subsection{{Adaptive h-DJSCC Transmission}}
In this subsection, we will first show the R-D performance of the proposed SNR-adaptive h-DJSCC transmission. Then, we evaluate the proposed fully adaptive h-DJSCC framework.

\begin{figure}[!t]
\centering
\includegraphics[width=0.8\columnwidth]{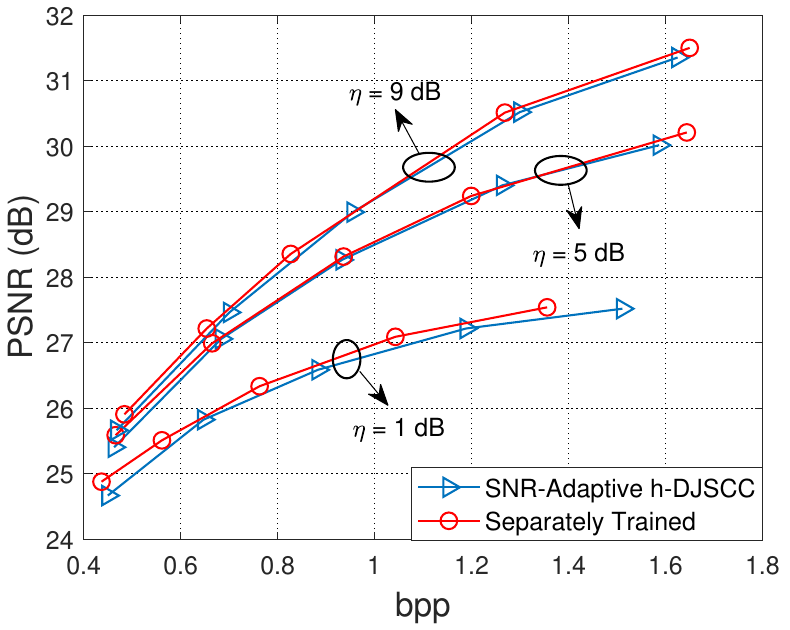}
\caption{Comparison of the R-D performance between the SNR-adaptive h-DJSCC models trained under varying $\eta \in [1, 9]$ dB values and the separately trained models.}
\label{fig:snr_adapt_vs_sep}
\end{figure}

\begin{figure}[!t]
\centering
\includegraphics[width=0.8\columnwidth]{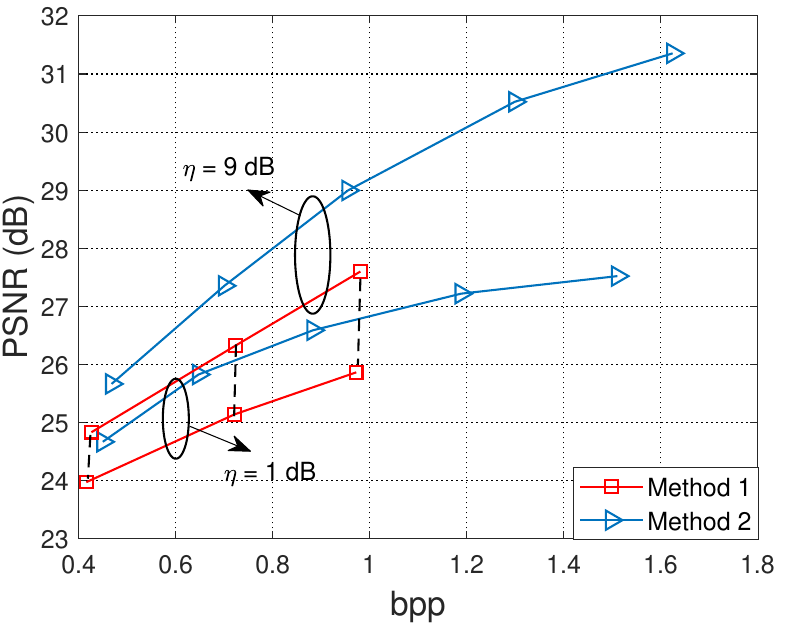}
\caption{R-D performance of the SNR-adaptive h-DJSCC model using different initialization methods.  The dashed lines indicate nearly identical bpp values w.r.t different $\eta$ values for the second method.}
\label{fig:train_method}
\end{figure}

\subsubsection{SNR-adaptive transmission}
In this case, we consider a scenario with varying SNR values $\eta \in [1, 9]$ dB and train five different SNR-adaptive models corresponding to $\lambda \in \{200, 400, 800, 1600, 3200\}$. As illustrated in Section \ref{sec:snr_adapt_jsc}, we use the pre-trained SNR-adaptive DeepJSCC model to initialize $f_s(\cdot, \eta)$ and $g_d(\cdot, \eta)$ of the SNR-adaptive h-DJSCC model. Note that for the separately trained models, even if the SNR values are chosen with an interval equals to 4 dB, i.e., $\eta = \{1, 5, 9\}$ dB, we still need to train 15 different models. 

The comparison of the R-D performance between the SNR-adaptive h-DJSCC models and the separately trained ones is shown in Fig. \ref{fig:snr_adapt_vs_sep}. For the SNR-adaptive h-DJSCC scheme, each R-D curve corresponds to a specific $\eta \in \{1, 5, 9\}$ value (with different $\lambda$ values). As can be seen, the SNR-adaptive h-DJSCC models can achieve nearly the same R-D performance with the separately trained models which manifests the effectiveness of the proposed SNR-aware neural networks and the underlying probability models in Section \ref{sec:snr_adapt_jsc}.

Then, we emphasize that the proposed initialization method is essential for satisfactory R-D performance. Two different initialization methods for the SNR-adaptive DeepJSCC encoder and decoder, i.e., $f_s(\cdot, \eta)$ and $g_d(\cdot, \eta)$, are evaluated. The first one (Method 1) adopts a random initialization while the second one (Method 2) loads the weights from the pre-trained SNR-adaptive DeepJSCC model and the weights are fixed during training\footnote{It is also possible to update/optimize the loaded weights during training, yet we find it leads to nearly the same performance compared with the fixed one.}. 

In this simulation, we train the SNR-adaptive h-DJSCC models corresponding to the two initialization methods under the same $\eta$ range and $\lambda$ values in Fig. \ref{fig:snr_adapt_vs_sep}. The relative performance of the two methods with $\eta = 1, 9$ dB is shown in Fig. \ref{fig:train_method}. We can find the naive initialization method\footnote{We show the results with $\lambda \in \{200, 1600, 3200\}$ for the random initialization method.} converges to sub-optimal solution whose R-D performance is significantly outperformed by the second method. It can also be seen that for a fixed $\lambda$, the bpp values of the random initialization method nearly keep the same w.r.t different $\eta$ values which matches the analysis of the empirical distribution, $\hat{P}_{\tilde{\bm{x}}_{\eta}}$ shown in Fig. \ref{fig:empirical_dist}. Thus, we adopt the second initialization method throughout this paper.

\begin{figure}[!t]
\centering
\includegraphics[width=0.8\columnwidth]{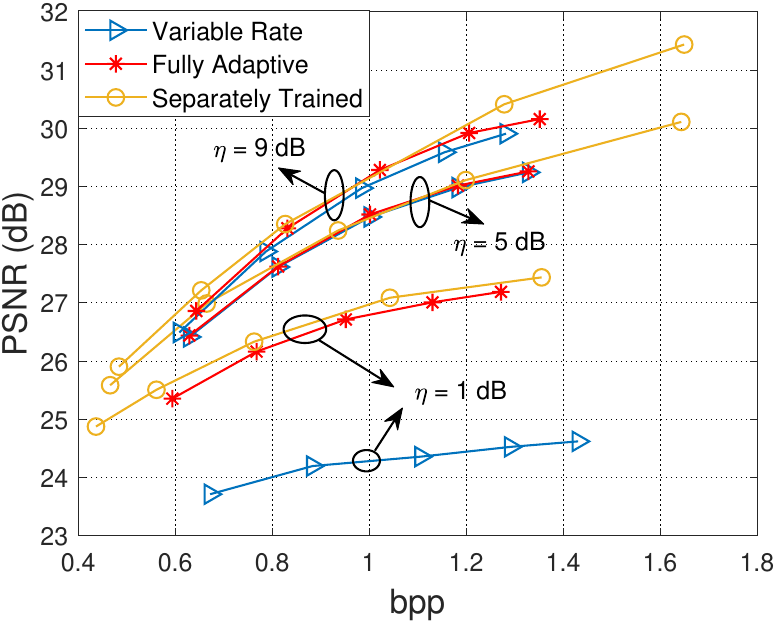}
\caption{{Comparison of the R-D performance between the fully adaptive model, the separately trained models and the variable rate h-DJSCC model trained at a fixed $\eta$ value.}}
\label{fig:full_adapt}
\end{figure}

\subsubsection{Fully adaptive h-DJSCC framework}
Then, we evaluate the fully adaptive h-DJSCC framework whose aim is to use a single model to provide satisfactory R-D performance for each and every combination of $\eta$ and $\lambda$ values. The model is trained with $\eta \in [1, 9]$ dB and $\lambda \in \{200, 400, 800, 1600, 3200\}$. 
  
The R-D performance of the fully adaptive h-DJSCC framework is compared with the separately trained models under different $\lambda$ and $\eta$ values and a variable rate h-DJSCC model trained at a fixed $\eta = 5$ dB (with the same $\lambda$ range). 
As shown in Fig. \ref{fig:full_adapt}, we evaluate the three models under $\eta = \{1, 5, 9\}$ dB. Though outperformed by the separately trained models, the fully adaptive model still achieves satisfactory R-D performance for each combination of $\eta$ and $\lambda$ values. When compared to the variable rate h-DJSCC model trained at a fixed $\eta = 5$ dB, we find the fully adaptive h-DJSCC model outperforms it when evaluated under $\eta = \{1, 9\}$ dB and achieves similar performance when $\eta = 5$ dB. 
This aligns with the intuition that, since no SA module is adopted for the variable rate h-DJSCC model, its DeepJSCC encoder outputs the same codeword regardless of the mismatched channel conditions leading to sub-optimal R-D performance.

\begin{figure}
    \centering
\begin{subfigure}{\columnwidth}
    \centering
    \includegraphics[width=0.8\columnwidth]{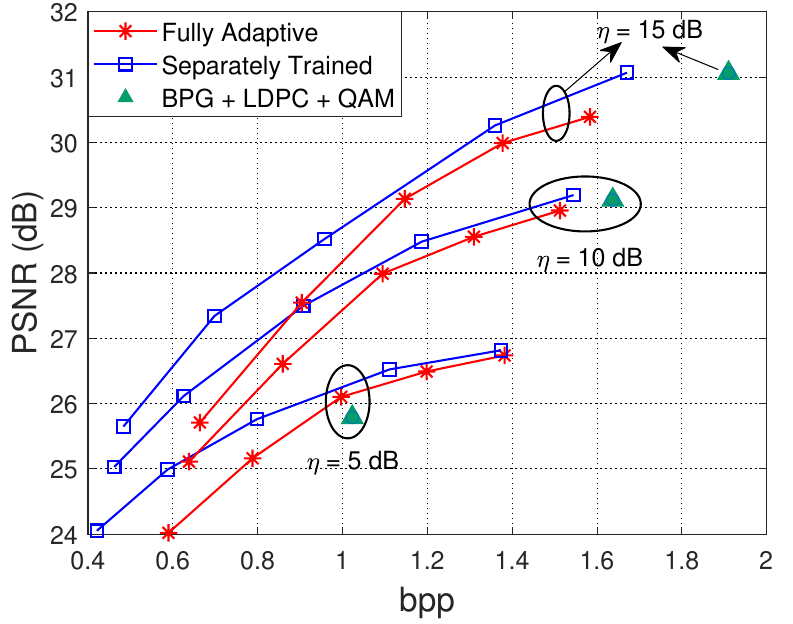}
    \caption{{CSI is available at both $\mathrm{S}$ and $\mathrm{R}_1$}.}
\end{subfigure}

\vspace{0.2cm}

\begin{subfigure}{\columnwidth}
    \centering
    \includegraphics[width=0.8\columnwidth]{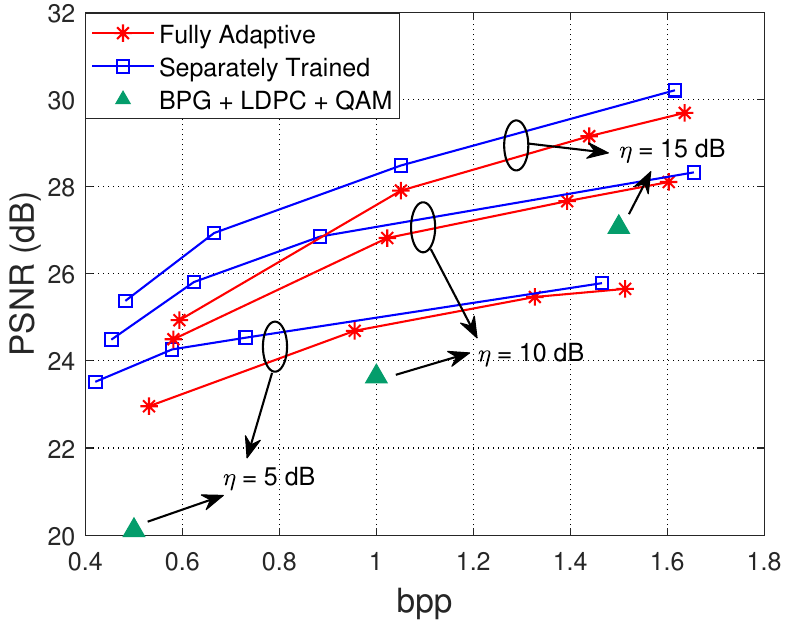}
    \caption{{CSI is only available at $\mathrm{R}_1$}.}
\end{subfigure}

\caption{{Comparison of the R-D performance between the fully adaptive model, the separately trained models and the digital baseline over Rayleigh fading channel under two different setups.}}
\label{fig:full_adapt_fading}
\end{figure}

\begin{figure}[!t]
\centering
\includegraphics[width=0.8\columnwidth]{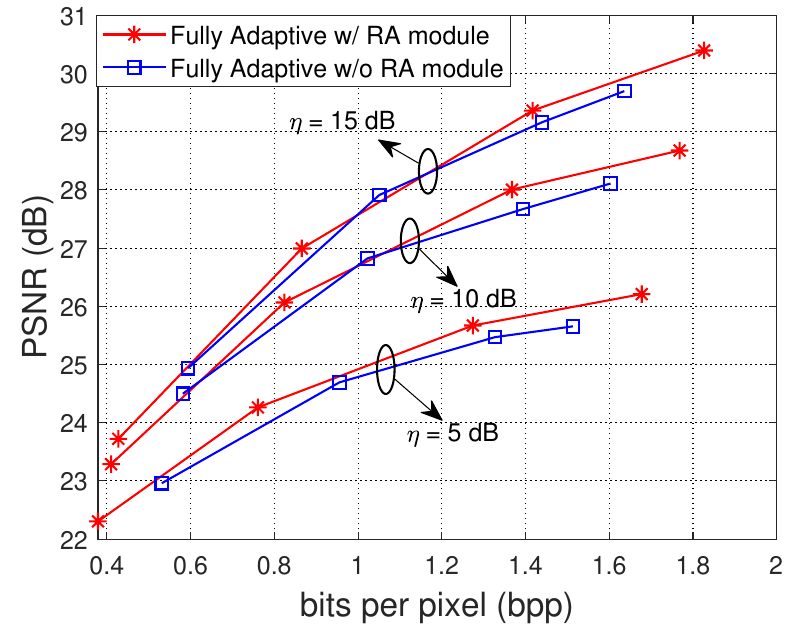}
\caption{{R-D performance of the fully adaptive h-DJSCC scheme with and without RA module under Rayleigh fading channel where the CSI is only available at $\mathrm{R}_1$.}}
\label{fig:ra_csir}
\end{figure}

\subsubsection{Fading channel}
To outline the effectiveness of the proposed h-DJSCC and the fully adaptive framework in the fading scenarios described in Section \ref{sec:fading}, we compare separately trained h-DJSCC models, the fully adaptive h-DJSCC model with the digital baseline which utilizes BPG compression algorithm and delivers the compression output using a combination of different coded modulation schemes which is analogous to the setup in Fig. \ref{fig:fig_rdtradeoff_w_cliff}. 
For both fading scenarios with and without CSIT, the fully adaptive model adopts the second initialization method in Fig. \ref{fig:train_method} and is trained with $\eta \sim \mathcal{U}(5, 15)$ dB and $\lambda \in \{200, 400, 800, 1600, 3200\}$. The separately trained models are trained under a fixed $\eta\in \{5, 10, 15\}$ value and evaluated at the same SNR. 
{We then briefly introduce the implementation of the digital baseline. For the fading scenario with CSIT, we select a specific LDPC code rate with a modulation order which exhibits sufficiently low BLER for each $\eta$ value and $h$ implementation, the corresponding rate is denoted as $R_{\eta, h}$. Note that zero BLER is impractical as the LDPC code considered in the paper exhibits error floor even in the high SNR regime. Since we calculate the average reconstruction performance over both successfully decoded images and the failed ones, selecting a proper coded modulation scheme which exhibits a low BLER is enough to guarantee satisfactory PSNR and SSIM performance. In our experiment, we set the BLER threshold to $10^{-3}$ and report the average PSNR performance for all the simulations.}
For the second scenario without CSIT, the transmitter has no access to $h$ and a fixed coded modulation scheme is adopted for each $\eta$ value. This leads to a non-zero BLER especially when the magnitude of $h$ is small. The PSNR performance is obtained by averaging over the successful decoding cases and the failed ones.

The performance of the two fading scenarios with and without CSIT are shown in Fig. \ref{fig:full_adapt_fading} (a) and (b), respectively.
For both scenarios, the fully adaptive model can achieve comparable R-D performance with the separately trained models with large $\lambda$ values, however, when $\lambda$ is small, e.g., $\lambda = 200$, there is a bpp gap around 0.2.
This aligns with the intuition that, it is hard to train a single model to be optimal for all the R-D points ($\lambda$ values) simultaneously. 
We also observe that for both scenarios with and without CSIT, the proposed h-DJSCC schemes outperform the digital baseline as the R-D points obtained by the baseline scheme are strictly below the curves achieved by the proposed schemes. A larger gain over the baseline is observed for the fading scenario without CSIT, which is due to the fact that the performance of the baseline scheme drops significantly when it is incapable to select different coded modulation schemes according to different channel realizations, $h$.   Note that our proposed schemes avoid the cliff and leveling effects in the meantime.

{We then show the effectiveness of the proposed RA module by comparing the R-D performance of the fully adaptive h-DJSCC model to that without the RA module. Due to the page limit, we only show their relative R-D performance under Rayleigh fading channel where the CSI is only available at $\mathrm{R}_1$ in Fig. \ref{fig:ra_csir}. The simulation setup is identical to that in Fig. \ref{fig:full_adapt_fading} (b). As can be seen, the fully adaptive h-DJSCC model with RA module achieves superior R-D performance. It is also observed that the model with RA module achieves a wider bpp range making it more suitable for real applications. }

  \begin{figure*}[t]
       \centering
    \captionsetup[subfigure]{labelformat=empty}
    \captionsetup{font={normalsize}}
    \begin{subfigure}[b]{0.19\linewidth}
       \centering
       \caption{\textbf{Ground Truth}}
       \includegraphics[width=60pt]{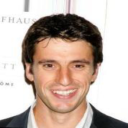}
       \vspace{-0.2cm}
       \caption{PSNR/SSIM/$L_{\bm{b}}$}
    \end{subfigure}%
    \begin{subfigure}[b]{0.19\linewidth}
       \centering
       \caption{\textbf{Fully Digital}}
       \includegraphics[width=60pt]{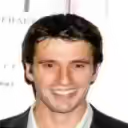}
       \vspace{-0.2cm}
       \caption{30.42/0.9280/6608}
    \end{subfigure}
    \begin{subfigure}[b]{0.19\linewidth}
       \centering
       \caption{\textbf{Fully Analog}}
       \includegraphics[width=60pt]{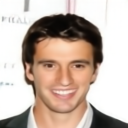}
       \vspace{-0.2cm}
       \caption{31.70/0.9377/nan}
    \end{subfigure}
    \begin{subfigure}[b]{0.19\linewidth}
       \centering
       \caption{\textbf{{\footnotesize Separately trained \\ h-DJSCC}}}
       \includegraphics[width=60pt]{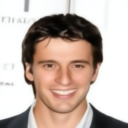}
       \vspace{-0.2cm}
       \caption{32.60/0.9467/4551}
    \end{subfigure}
    \begin{subfigure}[b]{0.19\linewidth}
       \centering
       \caption{\textbf{{\footnotesize Fully adaptive \\ \quad h-DJSCC}}}
       \includegraphics[width=60pt]{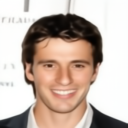}
       \vspace{-0.2cm}
       \caption{32.27/0.9433/4029}
    \end{subfigure}

    \captionsetup[subfigure]{labelformat=empty}
    \captionsetup{font={normalsize}}
    \begin{subfigure}[b]{0.19\linewidth}
       \centering
       \includegraphics[width=60pt]{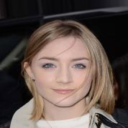}
       \vspace{-0.2cm}
       \caption{PSNR/SSIM/$L_{\bm{b}}$}
    \end{subfigure}%
    \begin{subfigure}[b]{0.19\linewidth}
       \centering
       \includegraphics[width=60pt]{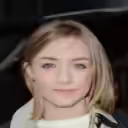}
       \vspace{-0.2cm}
       \caption{31.53/0.9230/5528}
    \end{subfigure}
    \begin{subfigure}[b]{0.19\linewidth}
       \centering
       \includegraphics[width=60pt]{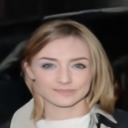}
       \vspace{-0.2cm}
       \caption{32.02/0.9187/nan}
    \end{subfigure}
    \begin{subfigure}[b]{0.19\linewidth}
       \centering
       \includegraphics[width=60pt]{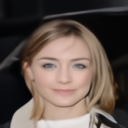}
       \vspace{-0.2cm}
       \caption{33.33/0.9326/4513}
    \end{subfigure}
    \begin{subfigure}[b]{0.19\linewidth}
       \centering
       \includegraphics[width=60pt]{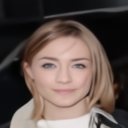}
       \vspace{-0.2cm}
       \caption{32.82/0.9261/4339}
    \end{subfigure}
    
    \captionsetup[subfigure]{labelformat=empty}
    \captionsetup{font={normalsize}}
    \begin{subfigure}[b]{0.19\linewidth}
       \centering
       \includegraphics[width=60pt]{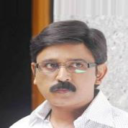}
       \vspace{-0.2cm}
       \caption{PSNR/SSIM/$L_{\bm{b}}$}
    \end{subfigure}%
    \begin{subfigure}[b]{0.19\linewidth}
       \centering
       \includegraphics[width=60pt]{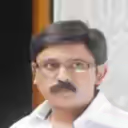}
       \vspace{-0.2cm}
       \caption{31.54/0.9231/5320}
    \end{subfigure}
    \begin{subfigure}[b]{0.19\linewidth}
       \centering
       \includegraphics[width=60pt]{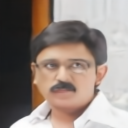}
       \vspace{-0.2cm}
       \caption{32.20/0.9086/nan}
    \end{subfigure}
    \begin{subfigure}[b]{0.19\linewidth}
       \centering
       \includegraphics[width=60pt]{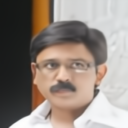}
       \vspace{-0.2cm}
       \caption{33.37/0.9215/4306}
    \end{subfigure}
    \begin{subfigure}[b]{0.19\linewidth}
       \centering
       \includegraphics[width=60pt]{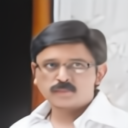}
       \vspace{-0.2cm}
       \caption{32.93/0.9185/4176}
    \end{subfigure}

\captionsetup{font={small}}   
\caption{Visual comparison between the reconstructed images from CelebA dataset obtained from the proposed h-DJSCC frameworks, the fully analog scheme and the fully digital baseline with $\eta = 5$ dB and the number of complex channel uses in the first hop is set to $k = 3072$. The PSNR, SSIM and {the number of bits ($L_{\bm{b}}$)} to be transmitted over the remaining hops are provided.}
\label{fig:fig_result_celeba}
\end{figure*}

\begin{figure}[!t]
\centering
\includegraphics[width=0.8\columnwidth]{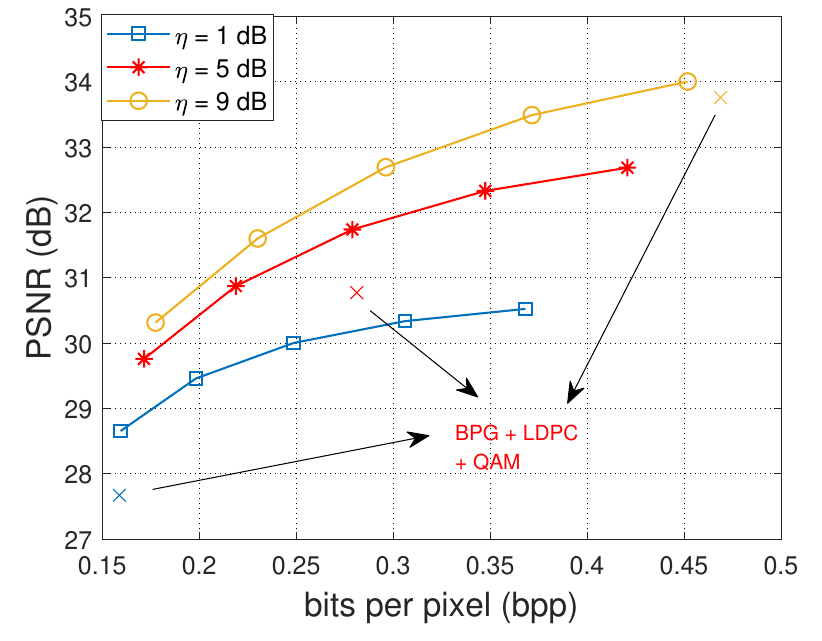}
\caption{{The relative R-D performance of the fully adaptive h-DJSCC model and the digital baseline evaluated under the CelebA dataset.}}
\label{fig:celeba_sim}
\end{figure}

\begin{table}[tbp]
\caption{The comparison of the storage used by different schemes, namely, 1. separately trained models, 2. SNR-adaptive models and 3. fully adaptive model.}
\begin{center}
\setlength{\tabcolsep}{3mm}
\begin{tabular}{c|ccc}
\hline

\cline{1-4} 
No. & 1 & 2 & 3\\
\hline

{Storage (Mb)} & 15*247.8  & 5*257.0& 257.1\\

\hline
\end{tabular}
\label{tab:storage}
\end{center}

\end{table}

\begin{table*}[t]
\begin{algocolor}
\caption{{The computational and time complexities among different nodes.}}
\begin{center}
\begin{tabular}{c|c|c|c}
\hline

 & \textbf{Source} ($\mathrm{S}$)& \textbf{Relay} ($\mathrm{R}_1$) & \textbf{Destination} ($\mathrm{D}$) \\
\hline

\textit{Computational complexity} & $\mathcal{O}(N_e C^2_{feat} K^2  H  W)$  & \makecell{$\mathcal{O}((N_d + N_l+ N_a)$\\ $C^2_{feat} K^2  H  W)$}  & \makecell{$\mathcal{O}((N_l+ N_a)$\\ $C^2_{feat} K^2  H  W)$}\\
\hline
\textit{Time complexity (s)} & 0.004 & 0.013 & 0.01\\

\hline
\end{tabular}
\label{tab:complexity}
\end{center}
\end{algocolor}
\end{table*}

\subsubsection{Reduction in storage}
It is worth mentioning the storage savings by adopting the fully adaptive h-DJSCC framework. As illustrated in Section \ref{sec:JSC}, the h-DJSCC framework can be divided into two parts, the DeepJSCC encoder ($f_s(\cdot)$) and decoder ($g_d(\cdot)$) as well as the compression ($f_c(\cdot)$) and decompression ($g_c(\cdot)$) modules. Denote the storage for $f_s(\cdot)$ and $g_d(\cdot)$ as $T_1$ while that for $f_c(\cdot)$ and $g_c(\cdot)$ as $T_2$, then for different $\lambda$ and $\eta$ values, a total amount of $ |\Lambda| |\mathrm{H}| (T_1 + T_2)$ space is required. The proposed fully adaptive h-DJSCC framework, on the other hand, only requires 
\begin{equation}
    T_1 + T_2 + \Delta T + 2c|\Lambda| (C_z + C_v)
\end{equation}
amount of storage where $\Delta T$ denotes the additional cost to store the SA modules while $c$ is a constant indicating the number of bytes to store each element of the scaling factors, $\{\bm{a}_\ell, \bm{a}^\prime_\ell, \bm{b}_\ell, \bm{b}^\prime_\ell\}, \ell\in [1, |\Lambda|]$. Since the additional terms are neglectable compared with $T_1$ and $T_2$, the total required storage can be significantly reduced.
In particular, we consider the following setting where $\lambda \in \{200, 400, 800, 1600, 3200\}$ and $\eta = \{1, 5, 9\}$ dB. The required storage of the (1) separately trained models, (2) the SNR-adaptive models and (3) the fully adaptive model are shown in Table. \ref{tab:storage}. As can be seen, the SA module and the scaling factors, $\{\bm{a}_\ell, \bm{a}^\prime_\ell, \bm{b}_\ell, \bm{b}^\prime_\ell\}, \ell\in [1, |\Lambda|]$ occupy merely $9.28$ and $0.02$ Mb, respectively yet adopting both of them is capable to reduce the overall storage cost by approximately $15\times$.

{  
\subsection{Complexity Analysis}
We then analyze both the computational and time complexities of the proposed h-DJSCC scheme for the source, $\mathrm{S}$, the relay, $\mathrm{R}_1$, and the destination node, $\mathrm{D}$.

To start with, the processing at the mobile user, $\mathrm{S}$, is simply a DeepJSCC encoder shown in Fig. \ref{fig:fig_NN} which consists of 2d CNN layers, SA and RA modules. In particular, the computational complexity for the CNN layer with $C_{feat}$ number of $K \times K$ kernels and unit stride can be expressed as: $\mathcal{O}(C_{in} C_{feat} K^2  H_{in}  W_{in})$ where $C_{in}, H_{in}, W_{in}$ denote the dimensions of the input tensor.  The complexity of the SA/RA module is $\mathcal{O}(C_{feat}^2)$. Then, the overall computational complexity is roughly $\mathcal{O}(N_e C^2_{feat} K^2  H  W)$, where $N_e$ is the number of CNN layers at the DeepJSCC encoder.

The relay node, $\mathrm{R}_1$, performs DeepJSCC decoding and image compression. The complexity of the DeepJSCC decoding mirrors that of the encoding which is expressed as $\mathcal{O}(N_d C^2_{feat} K^2  H  W)$, where $N_d$ is the number of CNN layers for the DeepJSCC decoder. The computational complexity of the image compression algorithm is dominated by the two non-linear transformations, $g_a, h_a$, whose complexities are given by $\mathcal{O}(N_l C^2_{feat} K^2  H  W )$ and $\mathcal{O}(N_a C^2_{feat} K^2  H  W)$, respectively, where $N_l$ and $N_a$ are the number of 2d CNN layers to produce the latent tensor, $\bm{z}$, and the hyper latent, $\bm{v}$, respectively. 
The arithmetic encoding is employed to generate the bit sequence from the quantized latents, $\hat{\bm{z}}$ of length-$C_z HW/16$ and $\hat{\bm{v}}$ of length-$C_v HW/256$ whose complexity can be expressed as $$\mathcal{O}(\log_2(|\mathcal{Z}|)C_z HW +\log_2(|\mathcal{V}|)C_v HW),$$
where $|\mathcal{Z}|$ and $|\mathcal{V}|$ represent the cardinality of the support sets for the latent vectors, $\hat{\bm{z}}$ and $\hat{\bm{v}}$, respectively. We find by experiments that the cardinality of the support sets, $\mathcal{Z}$ and $\mathcal{V}$ are smaller than $100$. Thus, the computational complexity of the arithmetic en/decoding processes can be omitted.

The destination node, $\mathrm{D}$, performs image decompression, whose computational complexity is the same with the image compression module. We summarize the computational complexity of each node in Table \ref{tab:complexity}.
Moreover, the run time complexity of the nodes are evaluated on a RTX 4080 GPU which is shown in Table \ref{tab:complexity}. As can be seen, the proposed h-DJSCC scheme is efficient where each image can be processed within 0.03 second across all nodes.

Finally, we illustrate the feasibility of applying semantic communication and image compression algorithms to devices with limited computing resources \cite{real_implement_semantic, MobileCodec}. In particular, the authors in \cite{real_implement_semantic} and \cite{MobileCodec} perform system-level implementation and evaluation of the DeepJSCC and the image compression schemes adopted in our manuscript, respectively. These two references verify that the h-DJSCC scheme can be applied to mobile devices which are becoming more and more powerful these days. 
}

\subsection{Larger Dataset}
Finally, we evaluate our scheme on the CelebA dataset \cite{celeba} with a resolution equals to $128 \times 128$ to show the proposed h-DJSCC framework is effective for different datasets and is capable to provide visually pleasing reconstruction. In this simulation, the number of complex channel uses of the h-DJSCC framework in the first hop is set to $k = 3072$ and we consider an AWGN channel with $\eta = 5$ dB. This corresponds to a channel capacity equals to $2.05$ which implies the maximum bit budget for the fully digital baseline is $\#\bm{b}_s \approx 6298$. For the proposed h-DJSCC scheme, the same neural network architecture shown in Fig. \ref{fig:fig_NN} is adopted where the output channel of the DeepJSCC encoder is set to $C_{out} = 6$. 
The h-DJSCC model is trained  with $\lambda = 1200$ leading to a bit budget smaller than that of the fully digital baseline, i.e., $6298$. For the fully adaptive h-DJSCC model, we adopt the same training setting as in the CIFAR-10 setup where the model is trained with $\lambda\in \{200, 400, 800, 1600, 3200\}$ and $\eta\in [1, 9]$ dB while evaluated at $\lambda = 800$ and $\eta = 5$ dB.

The reconstructed images produced by our models are shown in Fig. \ref{fig:fig_result_celeba} where we further provide a fully analog baseline adopting DeepJSCC-AF protocol illustrated in \cite{hybrid_jscc} which has three additional hops with identical channel qualities equal to $10$ dB. As can be seen, the proposed h-DJSCC models outperform the fully digital baseline adopting BPG compression algorithm delivered at the AWGN channel capacity. It also outperforms the fully analog scheme which does not perform neural compression at $\mathrm{R}_1$. Moreover, it can be found that the fully adaptive h-DJSCC model achieves comparable performance with the separately trained one. Compared with the fully digital baseline, the proposed h-DJSCC frameworks not only yield superior or equivalent PSNR and SSIM values but also produce visually pleasing reconstructions using less or equal amount of bits.

{Finally, we compare the R-D performances of the fully-adaptive h-DJSCC framework and the digital baseline over the images from the entire CelebA test dataset which is shown in Fig. \ref{fig:celeba_sim}. The simulation setup is identical to that in Fig. \ref{fig:fig_result_celeba}. As can be seen, the proposed scheme outperforms the digital baseline for all $\eta$ values.}

\section{Conclusion}
This paper introduced a novel h-DJSCC framework tailored to enhance the reliability and efficiency of image transmission across hybrid wireless and wired multi-hop networks. Our study revealed that traditional approaches, which rely solely on fully digital transmissions, often fail to meet the demands of complex network architectures. To address these shortcomings, we developed a hybrid solution that leverages DeepJSCC for the initial wireless hop to the access point, followed by a neural network-based compression model that effectively transforms the DeepJSCC codeword into a digital bitstream, suitable for stable transmission over subsequent wired hops to the edge server.

The implications of our work extend far beyond mere technical achievement. By providing a robust and adaptive solution that can effectively manage the complexities of transmitting high-quality images across hybrid multi-hop network conditions, we unlock new possibilities for real-world applications. From remote healthcare, where reliable and timely image transmission can be life-saving, to disaster response scenarios, where communication infrastructure is often compromised, our hybrid JSCC framework stands to revolutionize the way we handle data transmission in critical, bandwidth-sensitive environments.

\bibliographystyle{IEEEbib}
\bibliography{refs}

\begin{IEEEbiography}[{\includegraphics[width=1.1in,height=1.3in,clip,keepaspectratio]{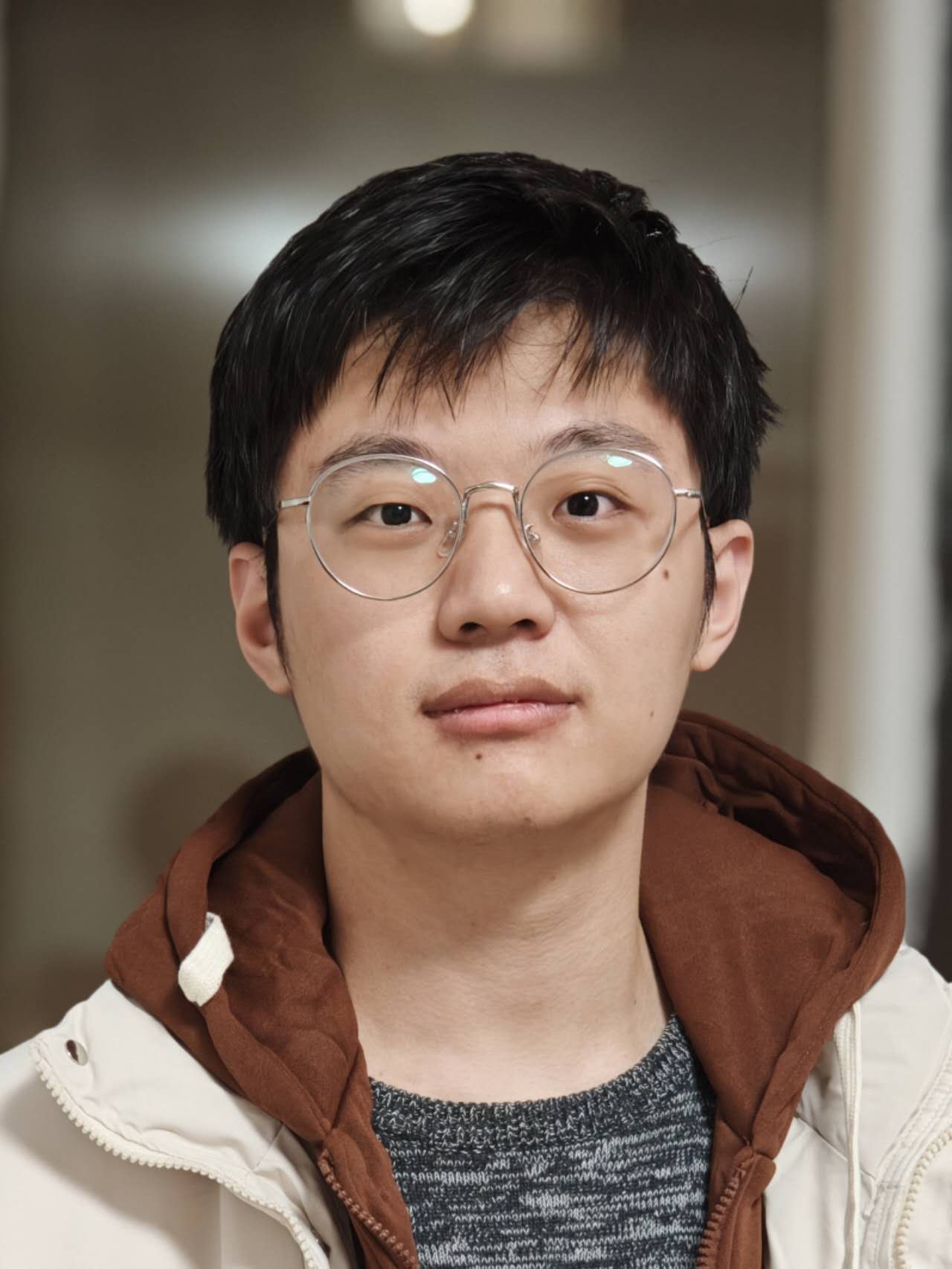}}]{Chenghong Bian} received the B.S. degree from the Physics Department, Tsinghua University, in 2020, and the M.S. degree from the EECS Department, University of Michigan, in 2022. He is currently pursuing the Ph.D. degree with the Department of Electrical and Electronic Engineering, Imperial College London. His research interests include wireless communications, machine learning and sensing.  He received the Best Paper Award at IEEE International Conference on Communications (ICC) 2023.
\end{IEEEbiography}

\begin{IEEEbiography}[{\includegraphics[width=1.1in,height=1.3in,clip,keepaspectratio]{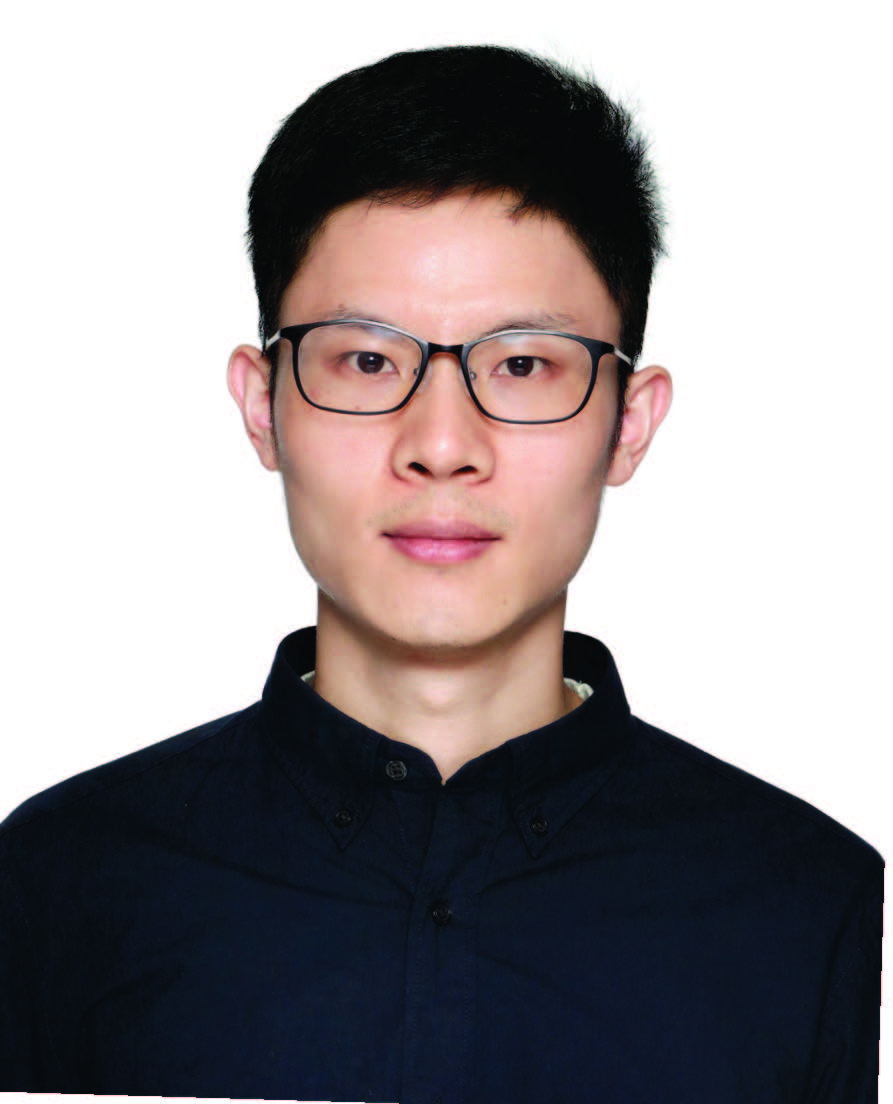}}]{Yulin Shao} (Member, IEEE) is an Assistant Professor with the State Key Laboratory of Internet of Things for Smart City, University of Macau, and a Visiting Researcher with the Department of Electrical and Electronic Engineering, Imperial College London. He received the B.S. and M.S. degrees in Communications and Information Engineering (Hons.) from Xidian University, China, in 2013 and 2016, and the Ph.D. degree in Information Engineering from the Chinese University of Hong Kong in 2020. He was a Research Assistant with the Institute of Network Coding, a Visiting Scholar with the Research Laboratory of Electronics at Massachusetts Institute of Technology, a Research Associate with the Department of Electrical and Electronic Engineering at Imperial College London, and a Lecturer in Information Processing with the University of Exeter. He was a Guest Lecturer at 5G Academy Italy and IEEE Information Theory Society Bangalore Chapter.

Dr. Shao's research interests include coding and modulation, machine learning, and stochastic control. He is a Series Editor of IEEE Communications Magazine in the area of Artificial Intelligence and Data Science for Communications, an Editor of IEEE Transactions on Communications in the area of Machine Learning and Communications, and an Editor of IEEE Communications Letters. He received the Best Paper Awards at IEEE International Conference on Communications (ICC) 2023, and IEEE Wireless Communications and Networking Conference (WCNC) 2024.
\end{IEEEbiography}

\begin{IEEEbiography}[{\includegraphics[width=1.1in,height=1.3in,clip,keepaspectratio]{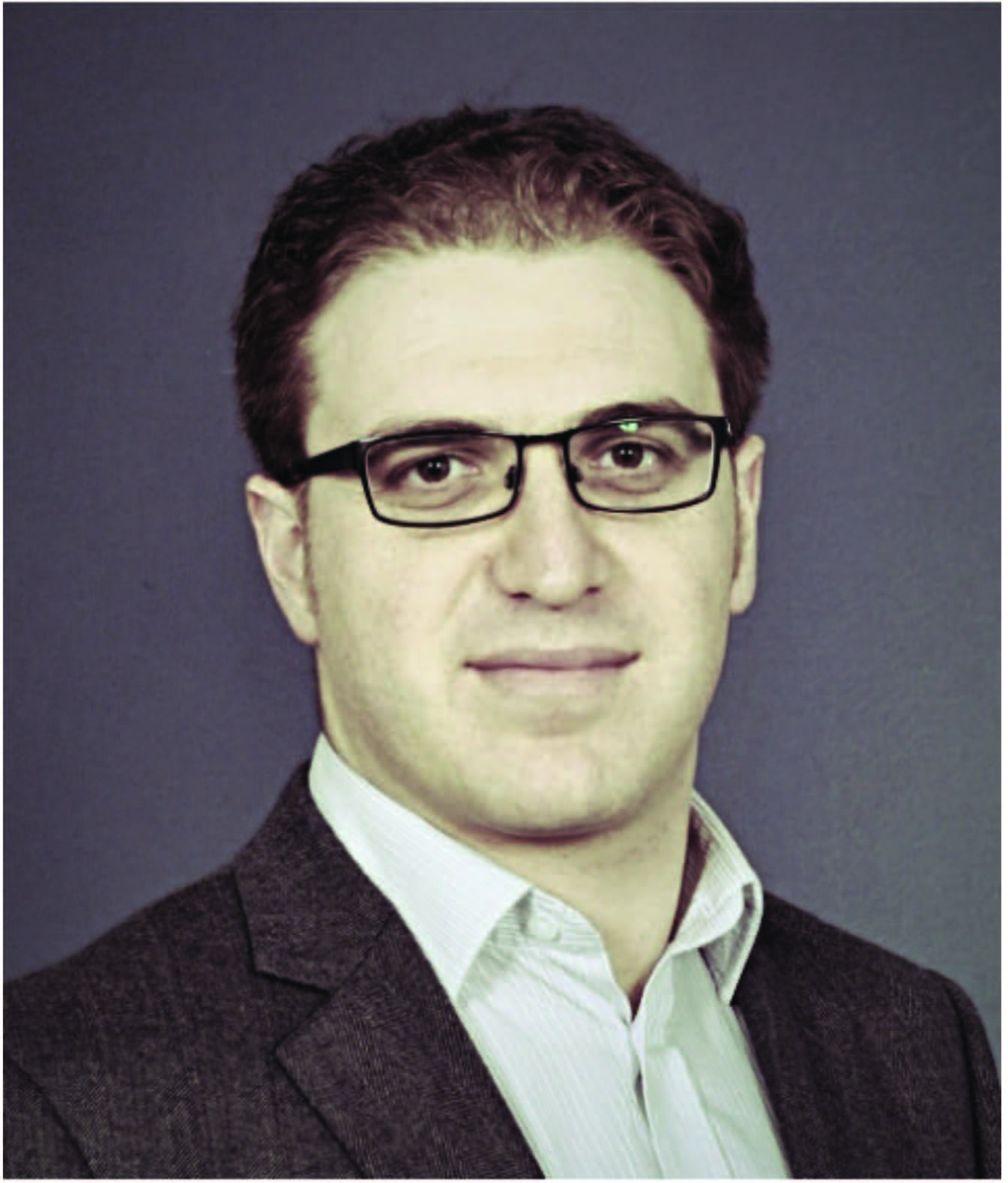}}]{Deniz~G\"und\"uz} (Fellow, IEEE) received the B.S. degree in electrical and electronics engineering from METU, Turkey in 2002, and the M.S. and Ph.D. degrees in electrical engineering from NYU Tandon School of Engineering (formerly Polytechnic University) in 2004 and 2007, respectively. Currently, he is a Professor of Information Processing in the Electrical and Electronic Engineering Department at Imperial College London, UK, where he also serves as the deputy head of the Intelligent Systems and Networks Group. In the past, he held various positions at the University of Modena and Reggio Emilia (part-time faculty member, 2019-22), University of Padova (visiting professor, 2018, 2020), Centre Tecnologic de Telecomunicacions de Catalunya (CTTC) (research associate, 2009-12), Princeton University (postdoctoral researcher, 2007-09, visiting researcher, 2009-11) and Stanford University (research assistant professor, 2007-09). His research interests lie in the areas of communications and information theory, machine learning, and privacy.

Deniz~G\"und\"uz is a Fellow of the IEEE. He is an elected member of the IEEE Signal Processing Society Signal Processing for Communications and Networking (SPCOM) and Machine Learning for Signal Processing (MLSP) Technical Committees. He serves as an Area Editor for the IEEE Transactions on Information Theory and IEEE Transactions on Communications. He is the recipient of the IEEE Communications Society Communication Theory Technical Committee (CTTC) Early Achievement Award in 2017, Starting (2016) and Consolidator (2022) and Proof-of-Concept (2023) Grants of the European Research Council (ERC), and has co-authored several award-winning papers, including the IEEE Communications Society - Young Author Best Paper Award (2022), and IEEE International Conference on Communications Best Paper Award (2023). He received the Imperial College London - President’s Award for Excellence in Research Supervision in 2023.
\end{IEEEbiography}

\end{document}